\documentclass[sigconf,nonacm,balance=false]{acmart}
\usepackage{popets}

\usepackage{hhline}
\usepackage{xspace}

\usepackage{subcaption}
\usepackage{tikz}
\usetikzlibrary{positioning, shapes.multipart, arrows.meta, decorations.pathmorphing}

\usepackage[shortlabels]{enumitem}
\usepackage[margin, draft]{fixme}
\fxsetup{theme=color,mode=multiuser}
\FXRegisterAuthor{gp}{agp}{GP}

\newcommand{\nopoint}[1]{\par\smallskip{\noindent\textbf{#1}~}}
\newcommand{\point}[1]{\par\smallskip{\noindent\textbf{#1.}~}}

\newlist{questions}{enumerate}{2}
\setlist[questions,1]{%
    label=\textbf{RQ\arabic*}.,
    ref=\textbf{RQ\arabic*},
    left=2pt,
    itemindent=0pt,
    topsep=2pt, 
    partopsep=0pt,
    parsep=0pt,
    itemsep=1pt
}
\setlist[questions,2]{%
    label=(\alph*),
    ref=\thequestionsi(\alph*),
    left=2em,
    itemindent=0pt,
    topsep=0pt,
    partopsep=0pt,
    parsep=0pt,
    itemsep=0pt
}

\newtheorem{theorem}{Theorem}
\newtheorem{remark}[theorem]{Remark}
\newcommand{\hwadversary}{$\mathsf{Adv}_\mathit{HW}$\xspace}
\newcommand{\swadversary}{$\mathsf{Adv}_\mathit{SW}$\xspace}

\newcommand{\connect}{$\boldsymbol{\mathsf{Connect(\mathnormal{n_1}, \mathnormal{n_2})}}$\xspace}
\newcommand{\verify}{$\boldsymbol{\mathsf{Verify(\mathnormal{n_1}, \mathnormal{n_2})}}$\xspace}
\newcommand{\sync}{$\boldsymbol{\mathsf{Sync(\mathnormal{n_1}, \mathnormal{n_2})}}$\xspace}
\newcommand{\attest}{$\boldsymbol{\mathsf{Attest(\mathnormal{n_1}, \mathnormal{n_2})}}$\xspace}
\newcommand{\getkey}{$\boldsymbol{\mathsf{GetKey(\mathnormal{ID}, \mathit{epoch})}}$\xspace}

\newcommand{\connectblank}{$\mathsf{Connect()}$\xspace}
\newcommand{\verifyblank}{$\mathsf{Verify()}$\xspace}
\newcommand{\syncblank}{$\mathsf{Sync()}$\xspace}
\newcommand{\attestblank}{$\mathsf{Attest()}$\xspace}
\newcommand{\getkeyblank}{$\mathsf{GetKey()}$\xspace}

\begin{document}

\title[Careful Whisper: Attestation for peer-to-peer Confidential Computing networks]{Careful Whisper: Attestation for peer-to-peer Confidential Computing networks}

\author{Ceren Kocaoğullar}
\affiliation{%
  \institution{University of Cambridge}
  \country{}}
\authornote{Corresponding author: ceren.kocaogullar@cl.cam.ac.uk}

\author{Gustavo Petri}
\affiliation{%
  \institution{Amazon Web Services}
  \country{}
}
\authornote{Work done whilst at Arm.}

\author{Dominic P. Mulligan}
\affiliation{%
  \institution{Amazon Web Services}
  \country{}
}
\authornotemark[2]

\author{Derek Miller}
\affiliation{%
 \institution{Arm}
 \country{}
}

\author{Hugo J. M. Vincent}
\affiliation{%
  \institution{Arm}
  \country{}
}

\author{Shale Xiong}
\affiliation{%
  \institution{Arm}
  \country{}
}

\author{Alastair R. Beresford}
\affiliation{%
  \institution{University of Cambridge}
  \country{}
}

\renewcommand{\shortauthors}{Kocaoğullar et al.}

\begin{abstract}
Trusted Execution Environments (TEEs) are designed to protect the privacy and integrity of data in use.
They enable secure data processing and sharing in peer-to-peer networks, such as vehicular ad hoc networks of autonomous vehicles, without compromising confidentiality.
In these networks, nodes must establish mutual trust to collaborate securely. 
TEEs can achieve this through remote attestation, where a prover presents evidence of its trustworthiness to a verifier, which then decides whether or not to trust the prover.
However, a naive peer-to-peer attestation approach, where every TEE directly attests every other TEE, results in quadratic communication overhead.
This is inefficient in dynamic environments, where nodes frequently join and leave the network.

To address this, we present Careful Whisper,  a gossip-based protocol that disseminates trust efficiently, reducing attestation overhead to linear complexity under ideal conditions. 
It enables interoperability by enabling transitive trust across heterogeneous networks, and supports trust establishment with offline nodes via relayed attestations. 
Using a custom discrete-event simulator, we show that Careful Whisper propagates trust both faster and more widely than naive approaches across various network topologies. 
Our results demonstrate that our protocol is resource efficient, sending  $\sim21.5$ KiB and requiring 0.158 seconds per round in a 200-node network, and that our protocol is resilient to attestation failures across various network topologies. 

\end{abstract} 


\maketitle

\section{Introduction}\label{sec:intro}

As the adoption of mobile and IoT devices, cloud computing, industrial control systems, and ambient computing continues to increase, the secure deployment and maintenance of large, dynamic, self-organising peer-to-peer networks becomes increasingly critical.
These networks can perform a wide range of security-, privacy-, or safety-critical tasks, such as preventing collision in self-driving cars, processing private data in the cloud, and securely sharing sensor data among IoT devices.
In such peer-to-peer networks, it is crucial to ensure that the data is kept private at all times, including while in use.

Consider, for example, a network of self-driving cars that collaboratively train machine learning models to improve their contextual awareness in a specific area, using data collected from various vehicles.
In such a setting with traditional computing, a malicious node can learn sensitive information about the drivers using data from other nodes, such as their driving habits, where they live and work, and behavioural patterns.
Even more concerning, a malicious node might deliberately manipulate the network's contextual understanding to cause accidents.
Even if each node encrypts its data while storing and shares them with others through secure channels, this does not guarantee privacy during computation, since in traditional computing the data is decrypted and processed in plaintext. 
This vulnerability persists even in federated learning, where each node trains on its own local data while contributing to a global model. 
Despite appearing privacy-preserving, federated learning has been shown to leak sensitive information through model updates and gradients \cite{zhu2019deep, bai2024membership, hitaj2017deep}.
One solution to this problem is Confidential Computing, which, as defined by the Confidential Computing Consortium \cite{confidential2022technical}, uses trusted hardware called Trusted Execution Environments (TEEs) to divide software and establish trust boundaries \cite{confidential2022technical}.
It aims to protect data during processing by providing isolated execution environments.

Equipping nodes of a security-, privacy-, or safety-critical peer-to-peer network with TEEs allows each node to process data while keeping it private.
However, for the nodes to collaborate on critical tasks, they must be able to establish trust among themselves. 
Here, trust refers to the assurance that one machine has in another's secure operation and adherence to confidentiality and integrity standards enabled by TEEs.
TEEs can perform \textit{remote attestation} to achieve this by allowing a TEE to prove its authenticity, integrity, and runtime status to a verifier.
The verifier in a remote attestation scheme can be a TEE. Therefore, two TEEs can establish reciprocal trust in a peer-to-peer setting by engaging in mutual attestation, each acting as both prover and verifier.
This is a significant improvement over authenticating nodes using a traditional digital certificate scheme such as TLS, which verifies only the identity of the node.
Remote attestation introduces a deeper layer of trust by allowing a verifier to evaluate the authenticity, software integrity, and certain runtime properties of the prover.

However, if every node in the network attests every other node, the resulting quadratic cost quickly becomes both computationally unscalable and financially expensive, especially when attestation services come at a cost. 
This challenge is compounded in heterogeneous networks, such as those made up of smart cars from different manufacturers, where nodes may use different TEEs and incompatible attestation protocols. 
In such cases, peer-to-peer trust cannot be established universally. 
The challenge is further compounded in dynamic networks, where nodes frequently go offline, making it impossible to establish trust with unavailable peers.

To address these limitations, we propose Careful Whisper, a machine-to-machine trust establishment mechanism for peer-to-peer TEE networks.
Careful Whisper supports networks with intermittent connectivity and diverse TEEs.
To disseminate trust information, it uses gossiping \cite{demers1987epidemic}, where nodes periodically exchange information with a small set of peers.
This approach leverages the idea that trust can be transitive when established and sustained within a trusted computing base.
Gossiping avoids the need for routing, topology awareness, or restrictions on network changes. 
Unlike collective attestation protocols (\S\ref{sec:careful-related-work}), Careful Whisper supports \textit{ad hoc} trust decisions, eliminating the need for central authorities during deployment or maintenance.

Our contributions are as follows:
\begin{itemize}
    \item We propose a peer-to-peer attestation protocol named Careful Whisper that enables self-deploying, self-configuring, and self-maintaining peer-to-peer Confidential Computing networks, reducing attestation validations from quadratic to near-linear complexity (\S\ref{sec:protocol}).
    \item We extend this protocol to constrain even powerful adversaries capable of bypassing TEE protections (\S\ref{sec:extension-against-hardware-adv}).
    \item We provide a comprehensive evaluation of Careful Whisper using a custom discrete-event simulator, across Erdős--Rényi, Watts--Strogatz, Barabási--Albert, and complete network topologies. Our evaluation demonstrates rapid trust propagation, low communication overhead via Bloom filters, and strong robustness under high failure rates (\S\ref{sec:evaluation}).
\end{itemize}
\section{Background}

\point{Confidential Computing}
Secure communication protocols, such as TLS, are used to protect data \textit{in transit}.
They are built on top of our long-standing ability to secure data \textit{at rest} using encryption.
Protecting data \textit{in use}, i.e. while being processed, remains a challenge.
Confidential Computing addresses this challenge by using hardware-based Trusted Execution Environments (TEEs) built for secure and isolated computation.
Several TEE implementations have been developed by major hardware vendors, such as Intel SGX \cite{mckeen2013innovative} 
and TDX \cite{inteltdx},
AMD SEV-SNP \cite{amdsev}, and Arm CCA \cite{armcca}.
Commercial cloud platforms, including Google Cloud \cite{googlecloudcc}, 
Microsoft Azure \cite{azurecc}, 
and Amazon Web Services \cite{awsnitro} have made Confidential Computing widely available.

By using specialised hardware, Confidential Computing provides distinctive advantages \cite{mulligan2021confidential} over other privacy-preserving computation techniques like homomorphic encryption and secure multiparty computation. 
Specifically, TEEs provide confidentiality for both data and code in execution through techniques such as hardware isolation and memory encryption.
Moreover, TEEs protect the integrity of the data and code throughout execution \cite{confidential2022technical}.
As a result, Confidential Computing is a versatile tool that can be used independently or to complement other private computation methods, such as homomorphic encryption. 
It is an essential component in building a foundation for comprehensive private computation.

\point{Remote attestation}\label{sec:background-remote-attestation}
Remote attestation is a key feature of Confidential Computing, as defined by the Linux Foundation's Confidential Computing Consortium \cite{confidential2022terminology}.
It allows a \emph{prover} device to provide information about the authenticity, integrity, and certain runtime properties of a TEE to a \emph{verifier} by providing hardware-signed evidence.
This information, also known as the attestation evidence, contains relevant measurements of the Trusted Computing Base (TCB) \cite{confidential2022terminology}, which is the collection of firmware and software that the security of a system depends on \cite{latham1986orangebook}.
The signature on the evidence also establishes that the attested node is running on a specific hardware instance or type of hardware.
The verifier then compares the attestation evidence against a set of trusted reference measurements and establishes trust in the prover if the evidence is found valid.
If the attestation is successful, the verifier may provision secrets or establish encrypted sessions tied to the TEE's integrity.

Remote attestation protocols may be interactive, requiring the verifier to initiate by sending a challenge to the prover \cite{eldefrawy2012smart, nunes2019vrased}, or non-interactive, requiring the prover to initiate, for example by directly generating and sending the attestation report to the verifier \cite{ibrahim2017seed}.
Remote attestation protocols can leverage different hardware or software environments to ensure that the prover generates legitimate attestation evidence \cite{rats}.
The security guarantees that remote attestation protocols promise can vary, depending on their construction and the hardware/software environments used.

\point{Collective attestation}
Collective (or swarm) attestation protocols aim to allow a trusted verifier to verify the trustworthiness of a swarm of provers, in a more efficient way than attesting every individual device. 
These protocols focus on a centralised model for attesting multiple devices.

The assumption of most collective attestation protocols is that the network is deployed by a trusted party; and also that a single entity, often the deployer, is responsible for verifying and obtaining information about the trustworthiness of the overall network or individual nodes \cite{ambrosin2016sana, carpent2017lightweight, ibrahim2016darpa, kohnhauser2017scapi, kohnhauser2018salad}.
Even schemes that allow multiple verifiers assume that the network is deployed and operated by a single party, and that all verifiers have knowledge of the public keys of all nodes in the network \cite{kohnhauser2019pasta, ambrosin2016sana}.
As a result, these protocols are ill-suited for decentralised deployment and maintenance of large peer-to-peer TEE networks.
We compare Careful Whisper to some prominent collective attestation protocols in \S\ref{sec:careful-related-work}.
\section{System model and goals}\label{sec:system-goals}
Having identified the limitations of the naïve attestation approach, we define the problem we aim to solve (\S\ref{sec:problem-definition}), present the objectives of our proposed protocol (\S\ref{sec:system-goals}), describe the system model (\S\ref{sec:system-model}), and the security considerations (\S\ref{sec:security-considerations}) in this section.

\subsection{Problem definition}\label{sec:problem-definition}

We now turn to analysing the limitations of naïve peer-to-peer attestation in the context of dynamic and heterogeneous peer-to-peer networks.

\point{Naïve peer-to-peer attestation is not scalable}
Consider a set $N$ of nodes employing the naïve protocol described above.
With $|N|$ nodes in the network, there are ${|N|\choose2}$ node pairs, and in every pair, both nodes must prove their state and verify the other's.
This leads to a total of ${|N|\choose2}*2 = |N|^2 - |N|$ attestations, resulting in quadratic complexity $O(|N|^2)$.
As the number of nodes increases, the number of attestation interactions grows quadratically, quickly becoming a bottleneck.

This scalability issue is further exacerbated when remote attestation protocols require interaction with trusted third-party services.
For example, Intel SGX’s attestation mechanism involves communication with the Intel Attestation Service (IAS) \cite{johnson2016intel, anati2013innovative}, introducing additional latency and network overhead.
Even non-interactive remote attestation protocols are not immune to this problem, as they still require each pair of nodes to exchange at least one message \cite{ambrosin2018pads, ibrahim2017seed, carpent2018erasmus}.
Therefore, onboarding a new node with the naïve peer-to-peer attestation into the system requires it to individually attest to and be attested by every existing node, leading to an explosion in communication and verification costs as the network scales.

\point{Naïve peer-to-peer attestation does not support heterogeneous networks}
In real-world peer-to-peer networks it is common for nodes to be manufactured by different vendors or run different versions of software and hardware platforms.
This diversity can lead to incompatibilities between attestation protocols, making it impossible for some nodes to mutually attest each other.

The naïve peer-to-peer attestation approach fails to establish trust between nodes that do not implement the same remote attestation protocol.
This problem has not been resolved by the collective attestation literature either (see \S\ref{sec:careful-related-work}).
While some collective attestation schemes claim support for different integrity measurement mechanisms, they typically rely on non-interactive attestation protocols where the prover can generate attestation evidence without engaging directly with the verifier \cite{kohnhauser2018salad}.
However, many attestation protocols used by real-world TEEs, including Intel SGX, require the prover to interact with the verifier to create the attestation proof \cite{menetrey2022attestation}.
This creates a significant interoperability barrier in heterogeneous networks.

\point{Naïve peer-to-peer attestation is not resilient to unstable connectivity}
Another key limitation of the naïve approach is its assumption of stable and consistent network connectivity.
In dynamic environments, such as edge deployments or mobile \textit{ad hoc} networks, this assumption does not hold.

If a node cannot reach another node due to temporary disconnection or network instability, it cannot complete mutual attestation, and trust between the two cannot be established.
Moreover, some schemes require access to external services to validate attestation reports \cite{johnson2016intel, anati2013innovative}. 
If a verifier cannot connect to such a service, the attestation process fails.
This shortcoming makes naïve peer-to-peer attestation unsuitable for deployments where network conditions are unpredictable or infrastructure-independent trust is essential.

\subsection{System goals}\label{sec:subsection-goals}
To describe the objectives and core mechanism of our protocol, we start with establishing that attestation forms the basis of trust between nodes (see \S\ref{sec:security-considerations}), and that this trust is transitive.

\begin{remark}[Transitivity of Trust]\label{lemma:transitivity}
Let $\mathbf{T}$ be a binary trust relation over a set $N$ of TEE-enabled nodes in a peer-to-peer network. Then:

\[
\forall n_A, n_B, n_C \in N.\; (n_A \mathbf{T} n_B \land n_B \mathbf{T} n_C) \Rightarrow n_A \mathbf{T} n_C
\]

\end{remark}

In plain terms, if node $n_A$ trusts node $n_B$, and $n_B$ trusts node $n_C$, $n_A$ also trusts $n_C$.
To show that this is correct, consider any $n_A, n_B, n_C \in N$, where $n_A\mathbf{T}n_B$ and $n_B\mathbf{T}n_C$.
In other words, we presume that $n_A$ trusts $n_B$ and $n_B$ trusts $n_C$.
We need to show that $n_A\mathbf{T}n_C$, or $n_A$ trusts $n_C$.
Now, assuming that attestation establishes trust on the attested TEE, if $n_B$ has received an evidence of attestation from $n_C$, it can propagate this evidence to $n_A$ having attested $n_B$.
That is, if $n_A$ attests $n_B$, and $n_B$ has evidence that $n_C$ has been attested (either by $n_B$ itself, or recursively through this procedure), then $n_A$ can rely on $n_B$’s account of $n_C$ being attested, and it does not need to further attest B. 
Therefore, we see that $n_A\mathbf{T}n_C$, as required.

Next, we establish the efficiency goal of Careful Whisper by discussing the lower bound of complexity for an ideal peer-to-peer attestation protocol.

\begin{remark}[Optimal Validation Complexity]\label{remark:optimal-linear}
In a connected network of $|N|$ nodes, the best-case time complexity for validating attestations from all nodes in $N$ is $O(|N|)$.
\end{remark}

Each node needs to be attested by at least one node to be trusted through attestation.
Therefore, we cannot establish trust among a network of nodes with a better complexity then $O(|N|)$.
Following Remark \ref{lemma:transitivity}, we can achieve this ideal time as follows.
We add each of the nodes in $N$ to the peer-to-peer network one by one, where each added node attests itself to one arbitrary node, and validates the attestation of one arbitrary node. 
A new node is onboarded only after the new node's successful attestation is communicated to all other nodes in the network.
As each incoming node is involved in 2 attestation operations, end-to-end trust is established in the network with $2*|N|$, hence $O(|N|)$ attestations.

\begin{remark}[Heterogeneous Transitivity]\label{remark:heterogeneity}
Let $p$ and $q$ be the only attestation protocols supported by nodes in the network $N$, and let $P$ and $Q$ be the sets defined as:

\[
P = \{n \in N \mid n \text{ supports } p\}, \quad Q = \{n \in N \mid n \text{ supports } q\}
\]

Then, $\forall n_A \in P - Q.\; \forall n_B \in P \cap Q.\; \forall n_C \in Q - P.\;$ it holds that:

\[
(n_A \mathbf{T} n_B \land n_B \mathbf{T} n_C) \iff n_A \mathbf{T} n_C
\]

\end{remark}

That is, nodes that do not support a common attestation algorithm can still establish trust among them, if and only if there exists a node that shares a supported attestation algorithm with each node.
Consider node $n_A$ supporting attestation protocol $p$, $n_C$ supporting $q$, and $n_B$ supporting both $p$ and $q$. 
Through transitivity of trust (Remark \ref{lemma:transitivity}), we get $(n_A\mathbf{T}n_B \land n_B\mathbf{T}n_C \Rightarrow n_A\mathbf{T}n_C)$.
For the other direction, since $n_A$ and $n_C$ do not share a common supported attestation protocol, they cannot attest each other directly.
Given that we assume that a node needs to be attested by at least one node to be trusted, and $n_A$ trusts $n_C$, this trust must have been established transitively through $n_B$.

\begin{remark}[Transitive Trust under Attestation Failure]\label{remark:connectivity}
Let $t$ be a given point in time, and let $N$ be the set of TEE-enabled nodes. Suppose that node $n_C \in N$ cannot be attested at time $t$. Then, for all $n_A, n_B \in N$:

\[
(n_A \mathbf{T} n_B \land n_B \mathbf{T} n_C) \iff n_A \mathbf{T} n_C
\]
\end{remark}

In simpler terms, if node $n_C$ cannot be attested at a given time, the only way for node $n_A$ to establish trust in $n_C$ at that time is by obtaining trust information from a third node $n_B$, which itself trusts $n_C$.
The proof follows similarly to that of Remark \ref{remark:heterogeneity}.
By the transitivity of trust (Remark \ref{lemma:transitivity}), $n_A$ can derive trust in $n_C$ through $n_B$, provided that $n_A$ already trusts $n_B$ and $n_B$ trusts $n_C$.
Therefore, we have $(n_A\mathbf{T}n_B \land n_B\mathbf{T}n_C \Rightarrow n_A\mathbf{T}n_C)$.
For the other direction, since $n_A$ cannot directly attest $n_C$ directly at time $t$, $n_A$ must establish trust with $n_C$ transitively through $n_B$.

Building on the above discussion, we now list the three core goals of our protocol. 

\medskip

\textbf{Careful Whisper aims to:}
\begin{enumerate}[label=\textbf{\arabic*.}, leftmargin=1.4em]
    \item Achieve peer-to-peer attestation with time complexity between $O(|N|)$ and $O(|N|^2)$.
    \item Support heterogeneous networks in which nodes implement different attestation protocols.
    \item Establish trust in environments with intermittent or unreliable connectivity.
\end{enumerate}

\subsection{System model}\label{sec:system-model}
We focus on peer-to-peer networks with no central operator to deploy nodes with certain features or enforce attestation criteria.
We consider both `unpermissioned' and `permissioned' versions of such networks. 
In unpermissioned networks, any node can join or leave the network freely. 
In contrast, permissioned networks only allow nodes owned or operated by authorised parties to join. 
The permissioned setup may be desirable in real-world systems like self-driving car networks or cloud computing networks, where control over network participants is necessary. 
In permissioned networks, we assume that nodes have the knowledge of allowed manufacturers and their public keys for verifying digital certificates.

We assume that nodes may support any attestation protocol of their choosing.
Furthermore, we consider that the network is dynamic, allowing nodes to join and leave at any time. 
We also consider that nodes may experience intermittent connectivity to other nodes in the network or to external services.
We do not make any assumptions about the number of nodes in the network at any given time or the topology of the network.
However, we assume that any node willing to join the network has knowledge of the set of hash functions necessary to construct and query Bloom filters \cite{bloom1970space} used to reduce network overhead (see \S\ref{sec:preliminaries}), and to hash trusted node information.

\subsection{Security considerations}\label{sec:security-considerations}
\point{Trust assumptions}
We assume all nodes are equipped with Trusted Execution Environments (TEEs) that can generate attestation reports to confirm the status of their hardware and software. 
These reports can be verified directly by other devices or through an external attestation validation service. 
We view attestation as a sufficient way of establishing complete trust in the TEE being attested, assuming that the code executed by each TEE being free of vulnerabilities. 
In the case of permissioned networks, where only allowed parties can join (see \S\ref{sec:system-model}), we assume that all parties on the list of allowed parties are trustworthy and protect their private signing keys from unauthorised third parties.

\point{Adversary model} 
The base version of Careful Whisper assumes a software-only adversary, denoted \swadversary.
This adversary can compromise the software of any node in the network, allowing them to observe and modify the memory and execution state that is not protected by a TEE.
Moreover, \swadversary has complete control over the communication links between the nodes, and can eavesdrop on and interfere with all exchanged messages. 
This adversary model is common in remote attestation literature, however, some collective attestation protocols (see \S\ref{sec:careful-related-work}), such as SALAD \cite{kohnhauser2018salad}, SANA \cite{ambrosin2016sana}, and DARPA \cite{ibrahim2016darpa} aim to provide some protection against hardware adversaries.
In \S\ref{sec:extension-against-hardware-adv}, we discuss a protocol extension that provides a similar degree level of protection against hardware adversaries.

\point{Security goals}
Our protocol aims to prevent \swadversary from deceiving any honest or compromised node into believing an unattested node $n$ as trustworthy, regardless of whether the \swadversary controls $n$ or not.
As is common in collective attestation protocols (see \S\ref{sec:careful-related-work}) \cite{kohnhauser2018salad, abera2019diat}, we consider denial-of-service attacks to be out of scope of this protocol, as \swadversary has full control over the communication between nodes and can discard any messages.
Nevertheless, we aim to limit the impact of denial-of-service attacks to only targeted nodes in permissioned networks (see \S\ref{sec:additional-features}). 
\section{The Careful Whisper protocol}\label{sec:protocol}

In this section, we describe the building blocks of the Careful Whisper protocol (\S\ref{sec:preliminaries}), describe the base protocol (\S\ref{sec:protocol-description}), discuss some additional features that can be achieved by making some implementation decisions (\S\ref{sec:additional-features}), finally discussing some limitations (\S\ref{sec:limitations}).

\subsection{Preliminaries}\label{sec:preliminaries}
\point{Node IDs}
In order to ensure that each node in the network can be distinctly identified, it is necessary for each node to have a unique ID that is stored securely in the TEEs. 
One possible approach to generate IDs is to use a keyed pseudorandom function that is seeded by each TEE using its own public key or a key derived from it. 
This method ensures that clashes between node IDs are prevented without requiring coordination between nodes. 
Although this approach requires each node to know which keyed pseudorandom function to use, this information can be made publicly available, and only needs to be set up by the node owner once.
Alternatively, if the TEE already has a unique ID assigned by its manufacturer or owner, it can use that instead.

\point{Policies} 
As we discuss in \S\ref{sec:system-model}, we design our protocol for networks without a central operator that can control node deployment or enforce attestation requirements.
Furthermore, each node can support any attestation protocol of its choice.
This brings out the need for a mechanism to ensure that each node can establish trust transitively as discussed in \S\ref{sec:system-goals}, while verifying the exact properties that it would check for if it attested those nodes directly.
We propose a policy framework to achieve this.

We define a policy as a data structure that contains information about the criteria that a verifier has attested the prover against. 
A policy can also include additional information, such as an attestation timestamp or expiry period (see \S\ref{sec:additional-features}).
Upon completing an attestation, each verifier generates a policy for the prover and records it in its list of trusted nodes. 
When sharing lists of trusted nodes, nodes also share the recorded policies for each node in the list.
To ensure consistent interpretation of policies across heterogeneous nodes, i.e. nodes that can attest each other, a predefined set of rules must be established.
These rules specify how criteria are encoded, what values are valid, and how additional fields like timestamps are to be processed.
All new nodes of the same type joining the network are expected to follow these rules when creating or interpreting policies.
For example, such rules could be defined using a domain-specific language, or by specifying a clear mapping between criteria and their possible values.

\point{Trusted nodes list} 
Each node in the network should maintain a list of trusted nodes, where each entry in the list is a mapping of a node ID and a policy.
Trusted nodes lists may have multiple entries for the same node ID, each with different policies.
To ensure the security and integrity of the trusted nodes list, it should be stored in a secure manner.
This can be achieved by storing the list in the secure memory inside the TEE.
However, since the list size grows with the networks size and no assumptions are made about the secure memory requirements, the list may exceed the capacity of secure memory.
In such a case, the list may be encrypted with a key stored in the secure memory using an integrity-protected encryption method, and stored outside the TEE.

\point{Trusted node Bloom filters}
To be able to establish transitive trust relationships, nodes must be able to communicate their trusted nodes with other nodes.
However, exchanging full lists of node IDs can increase network overhead in large networks. 
To address this issue, nodes can use Bloom filters \cite{bloom1970space} instead of exchanging full lists.

A Bloom filter is a probabilistic data structure that is used for testing if an element is member of a set or not. 
It is a memory-efficient construct, consisting of a bit array of length~$m$, initially all set to~0. 
To insert an element, it is hashed~$k$ times with independent hash functions, and the resulting positions in the array are set to~1. 
To check membership, the same~$k$ hashes are computed; if all corresponding bits are set to~1, the element is possibly in the set, otherwise it is definitely not. 
In other words, a Bloom filter can give false positives, but not false negatives.
The false positive probability after inserting~$n$ elements is ~$\sim\left(1 - e^{-kn/m} \right)^k$.
Bloom filters do not inherently support deletion. 

In Careful Whisper, each node \( n_A \) shares a Bloom filter with \( n_B \) as a compact representation of its trusted node list.  
Node \( n_B \) can then use this Bloom filter to identify entries in its own trusted list that are likely not present in \( n_A \)'s list.  
With high probability, this allows \( n_B \) to share only the missing entries with \( n_A \), rather than its entire trusted node list.

\begin{figure}
    \centering
    \includegraphics[width=\linewidth]{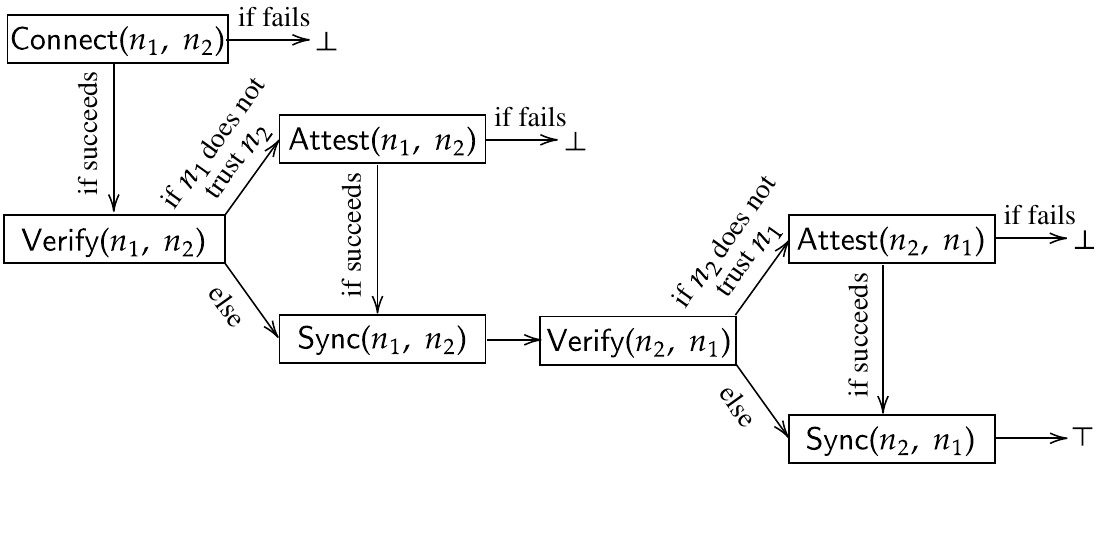}
    \caption{The diagram of Careful Whisper's subprotocols.}
\end{figure}\label{fig:protocol_flow}

\begin{figure*}[!ht]
    \centering
    \includegraphics[width=0.9\textwidth]{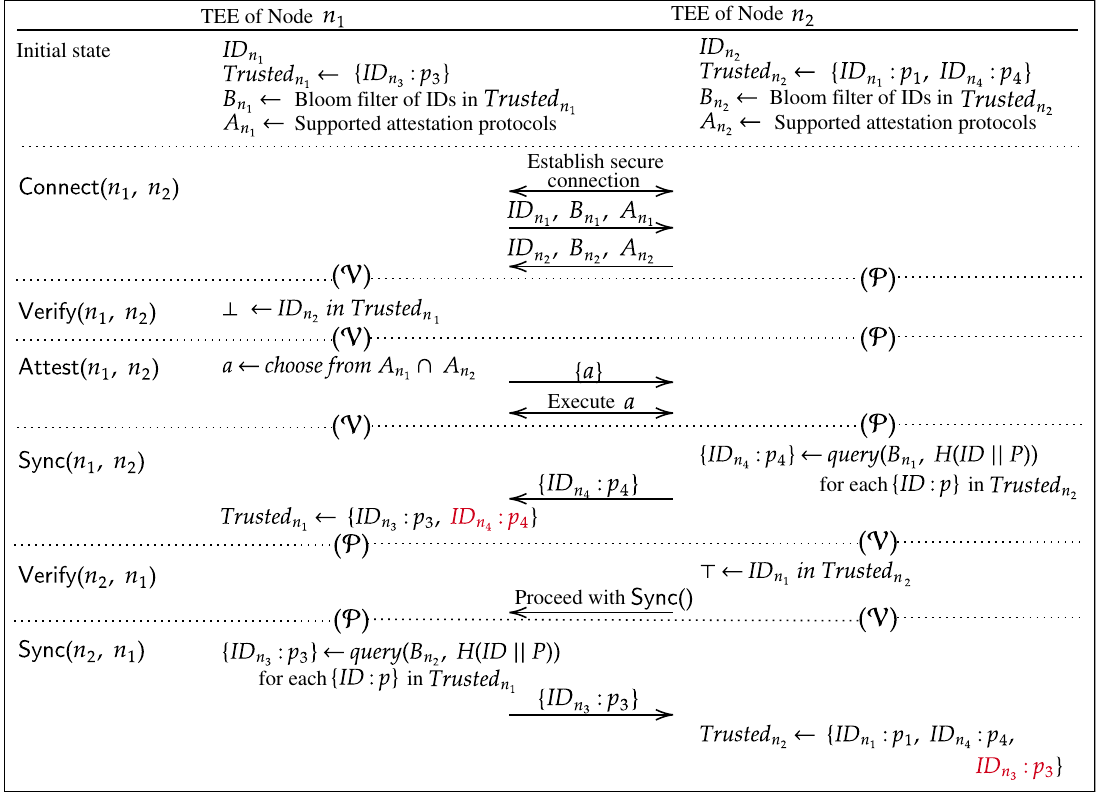}
    \caption{An example Careful Whisper protocol flow between nodes $n_1$ and $n_2$, where $n_1$ does not have prior trust on $n_2$, while $n_2$ already trusts $n_1$. The prover and verifier roles of the two nodes for \verifyblank, \syncblank, and \attestblank subprotocols are denoted as $\mathcal{P}$ and $\mathcal{V}$, respectively.}
    \label{fig:protocol_detail}
\end{figure*}

\subsection{Protocol overview}\label{sec:protocol-overview}

The Careful Whisper protocol consists of four subprotocols, namely \connectblank, \verifyblank, \syncblank, \attestblank, which we describe in detail later in this section.
These subprotocols together form the Careful Whisper protocol as follows, and as described in Figure \ref{fig:protocol_flow}.
\begin{enumerate}
    \item Initially, two nodes establish secure connection between their TEEs (\connectblank).
    \item One node assumes the role of the verifier, and checks if the other node, the prover is already in its list of trusted nodes (\verifyblank). \label{step:verify}
    \begin{enumerate}
        \item If the verifier is not yet trusted, the prover employs a remote attestation protocol to prove its trustworthiness to the verifier (\attestblank).\label{item:attest}
        \item If the prover is already trusted or if the attestation from Step \ref{item:attest} is successful, the verifier expands its list of trusted nodes with the entries from the prover's list (\syncblank).
    \end{enumerate}
    \item Once the one-sided trust establishment is complete, the roles of the nodes switch, and the process is repeated from Step \ref{step:verify}.
\end{enumerate}

\subsection{The subprotocols}\label{sec:protocol-description}
After outlining the Careful Whisper protocol flow, we describe the details of the subprotocols below.
Figure \ref{fig:protocol_detail} provides a schema showing the structure and execution sequence of the subprotocols.
The operations included in all these subprotocols occur inside the TEEs, and in our descriptions in this section, we use the word `node' to refer to the TEE.

\nopoint{\connect}
This subprotocol establishes a secure communication channel between two nodes and exchanges basic trust-related metadata.
\begin{enumerate}
    \item TEEs of the two nodes $n_1$ and $n_2$ establish a secure communication channel between each other. 
    For permissionless networks (see \S\ref{sec:system-model}), a non-authenticated key agreement protocol like Diffie-Hellman can be used for secure connection establishment. 
    On the other hand, for permissioned networks, an authenticated protocol such as TLS can be employed.
    \item Once the secure communication channel is established, each node sends a message to the other one with three pieces of information: (1) its ID, (2) a Bloom filter pre-populated with hashes of the node ID-policy mappings in its trusted nodes list, (3) the set of remote attestation protocols it supports. 
    \item Once this subprotocol succeeds, any one of the nodes initiates the \verifyblank protocol, and assumes the role of the initial prover. 
\end{enumerate}

\nopoint{\verify}
This subprotocol determines whether the prover ($n_2$) is already trusted by the verifier ($n_1$) and decides whether to proceed with attestation or synchronisation.

\begin{enumerate}
    \item $n_1$ checks its list of trusted nodes to determine whether it already trusts $n_2$.
    \begin{enumerate}
        \item If so, $n_1$ sends a message to $n_2$, signalling it to initiate \syncblank.
        \item Otherwise, $n_1$ initiates \attestblank.
    \end{enumerate}
\end{enumerate}

\nopoint{\sync}
This subprotocol allows the verifier ($n_1$) to expand its list of trusted nodes with the prover's ($n_2$) list of trusted nodes by selectively exchanging missing entries.

\begin{enumerate}
    \item $n_2$ queries the Bloom filter provided by $n_1$ in \connectblank subprotocol, using each of its trusted node ID - policy pairs.
    This allows $n_2$ to identify which entries are likely missing from $n_1$'s trusted nodes list. 
    \item \label{item:sync-send-list} $n_2$ then sends a message to the verifier containing a list of these potentially missing entries.
    \item \label{item:sync-update-list} $n_1$ updates its list of trusted nodes with the list of nodes it has received from $n_2$.
    If an entry with the same node ID already exists, the verifier may choose to retain only one version, based on factors such as memory constraints or preference for newer attestation timestamps.
    \item If $n_2$ has not acted as a verifier throughout the current Careful Whisper protocol flow, it runs $\mathsf{Verify(\mathnormal{n_2}, \mathnormal{n_1})}$. 
    This effectively switches the verifier and prover roles for the rest of the protocol flow.
    If both nodes have already acted as both prover and verifier, the protocol run for $n_1$ and $n_2$ is complete.
\end{enumerate}

\nopoint{\attest}
This subprotocol performs remote attestation between the prover ($n_2$) and verifier ($n_1$), and records a policy if successful.

\begin{enumerate}
    \item $n_1$ compares its supported attestation protocols with those of $n_2$, as received during \connectblank. 
    This step is necessary even for non-interactive protocols, since the verifier must inform the prover whether to proceed with attestation.
    \begin{enumerate}
        \item If the two nodes know at least one common attestation protocol, $n_1$ picks any protocol from the intersection set, and sends a message to $n_2$ to trigger the attestation protocol.
        If the chosen attestation protocol is interactive and requires any additional data, such as a challenge value, from the verifier, $n_1$ also provides this data in this message.  
        \begin{enumerate}
            \item \label{item:attest-succeeds} If the attestation succeeds, $n_1$ first creates a policy instance for $n_2$. 
            This policy includes information about what criteria $n_2$ was attested against.
            After this, $n_1$ runs $\mathsf{Sync(\mathnormal{n_1}, \mathnormal{n_2})}$.
            \item Otherwise, the Careful Whisper protocol flow is terminated.
            $n_1$'s trust establishment with $n_2$ fails.
        \end{enumerate}
        \item If the nodes do not know any common attestation protocols or the attestation protocol fails, both this subprotocol and Careful Whisper protocol terminates.
    \end{enumerate}
\end{enumerate}

\subsection{Revisiting the goals}
In this section, we assess Careful Whisper with respect to the protocol goals outlined in \S\ref{sec:system-goals}.

\medskip
\point{Goal 1: Careful Whisper aims to achieve peer-to-peer attestation in between $O(|N|)$ and $O(|N|^2)$ time}

\medskip
In a peer-to-peer network, nodes joining the network can do so at any time and in any order. 
Ensuring that all nodes join the network sequentially as described in Remark \ref{remark:optimal-linear} is not always feasible. 
As a result, a node may need to attest to multiple nodes if it does not receive an attestation report for a newly contacted node. 
This means that real-world peer-to-peer attestation cannot always achieve $O(|N|)$ time complexity. 
In the worst case scenario, every node needs to attest to every other node in the network, which is the case in the naïve protocol outlined in \S\ref{sec:problem-definition}. 
This can occur even in a protocol that uses transitivity of trust, particularly if the network is highly fragmented. 
Therefore, our protocol aims to operate in amortised time between $O(|N|)$ and $O(|N|^2)$, given these upper and lower bounds.

\point{Goal 2: Careful Whisper aims to accommodate networks with nodes that support various attestation protocols}

\medskip

Remark \ref{remark:heterogeneity} highlights that trust can be established between nodes that do not share a common attestation protocol by leveraging the transitivity of trust.
This is made possible through intermediary nodes that support multiple attestation protocols, effectively acting as bridges for trust propagation.
By enabling trust formation in heterogeneous networks, Careful Whisper aims to support collaborative systems involving diverse stakeholders, such as those found in smart cities or networks of self-driving vehicles from different manufacturers.
Moreover, the protocol’s ability to accommodate various attestation schemes promotes backwards compatibility and interoperability, making it easier to integrate legacy systems and heterogeneous devices into the network.
This is especially important in real-world deployments, where the use of specialised hardware for TEEs can make it impractical or costly to maintain uniformity across all nodes.

\medskip
\point{Goal 3: Careful Whisper aims to enable trust establishment in networks with unreliable connectivity}
\medskip

Consider a scenario where a node requires internet access to complete its attestation, but temporarily loses connectivity to external services essential to the protocol.
Alternatively, a node may become unreachable by its peers, preventing it from being attested altogether.
In such cases, it becomes difficult to establish trust in a timely or reliable manner.
As discussed in Remark \ref{remark:connectivity}, Careful Whisper addresses these challenges by allowing nodes to continue making informed trust decisions, even when another node cannot be attested within a given time window.
This goal is particularly relevant in environments with intermittent connectivity, such as delay-tolerant networks, or in highly dynamic systems where nodes frequently join and leave the network.

\begin{figure}[t]
    \centering
    \includegraphics[width=\linewidth]{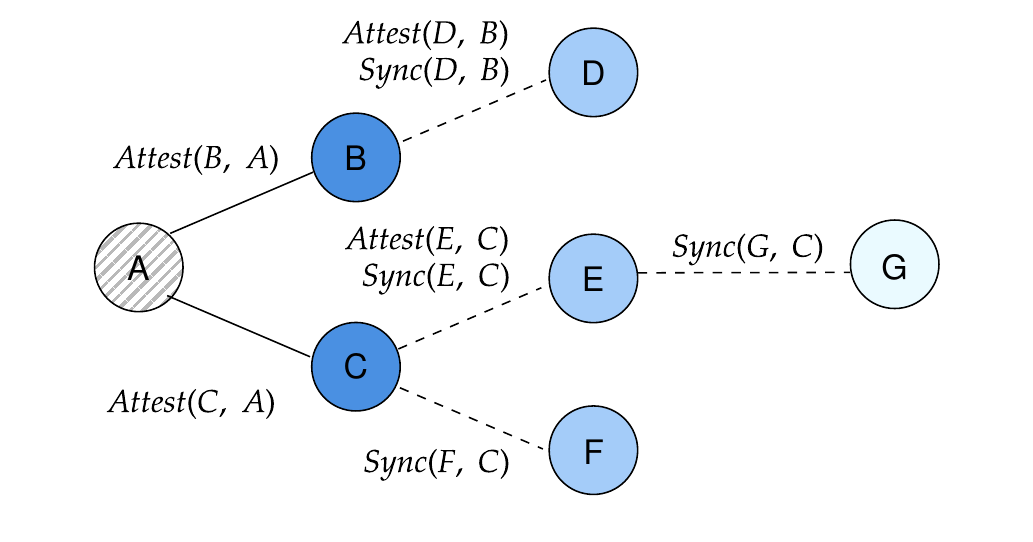}
    \caption{Trustworthiness of node~A propagating through a peer-to-peer network using Careful Whisper via multiple paths: 
[1] direct attestation of A (B and C); 
[2] transitive trust established by synchronising with an already trusted node (F and G); and 
[3] trust establishment with a previously untrusted node, followed by the inclusion of its trusted node A (D and E).
}
\end{figure}\label{fig:protocol_flow}

\subsection{Additional considerations}\label{sec:additional-features}

\point{Denial-of-Service resistance}
While our protocol does not guarantee perfect denial-of-service protection, it limits an attacker's ability to perform such attacks in permissioned networks (see \S\ref{sec:system-model}). 
In these networks, nodes can share certificates obtained from their owner or manufacturer in the \connectblank phase, allowing the verifier to confirm whether a node is authorised to join the network and blacklist it if necessary. 
As we assume software-only attackers who cannot modify the protocol code inside the TEE, this measure prevents the attacker from using the protocol to launch denial-of-service attacks against nodes that it does not control.
Despite this, our protocol cannot provide full denial-of-service protection since the attacker can still drop messages, as described in \S\ref{sec:security-considerations}.

\point{Removing nodes} 
Our protocol includes a mechanism to detect nodes that do not maintain the criteria they were attested against and revoke trust accordingly.
This is achieved by including an attestation timestamp or expiry date in the prover's policy, which is used by each verifier to decide when to expire a node from its trusted nodes list.
Upon receiving the list of node IDs and policies in \syncblank, each node checks if the attestation is still valid before adding each entry to its trusted nodes list.
Apart from this, the nodes should also keep track of their trusted node lists for any expired trust relationships.
Since it is not possible to remove any elements from Bloom filters, nodes need to repopulate their Bloom filter with the updated trusted nodes list upon removing a node.
If maintaining trust with an expired node is deemed urgent, the node that removed it from the trusted nodes list can immediately execute \attestblank subprotocol to reestablish trust.

\subsection{Limitations}\label{sec:limitations}
Our protocol makes no assumptions about the attestation protocols supported by the nodes in the network. 
Therefore, there may be nodes that cannot join the network because they do not support any of the attestation protocols used by other nodes.
Another limitation is that an attacker may flood a permissionless network with attestation attempts and launch a denial-of-service (DoS) attack.

Since Bloom filters may produce false positives \cite{bloom1970space}, the prover may incorrectly classify a node as already trusted by the verifier in \syncblank. 
This does not cause any security issues, but decreases the efficiency of the protocol, since each false positive will result in the verifier losing the opportunity to establish transitive trust with a node.
The likelihood of false positives increases with the number of entries in a trusted nodes list.
However, the probability of false positives can be decreased by choosing the number of hash functions used in the Bloom filter and the filter size appropriately \cite{bloom1970space, mullin1983second}.

Lastly, this protocol does not provide protection against powerful adversaries with hardware access.
We discuss a protocol extension to mitigate this issue in the next section.

\section{Protection against hardware adversary}\label{sec:extension-against-hardware-adv}
The Careful Whisper protocol described in Section~\ref{sec:protocol} assumes a software-only adversary and relies on the integrity of the TEEs used on the nodes, without accounting for potential hardware-level compromises (see \S\ref{sec:security-considerations}). 
However, a stronger adversary with physical access may be able to bypass TEE protections by exploiting hardware vulnerabilities, such as side-channel attacks or fault injection~\cite{trippel2017clkscrew, van2018foreshadow}, to forge a ``trusted nodes list'' that includes malicious nodes, causing honest nodes to trust them based on transitivity of trust.

In this section, we introduce an extension to the base Careful Whisper protocol (Section~\ref{sec:protocol}) designed to significantly restrict the impact of such physical adversaries in permissioned networks. 
This enhancement trades off increased memory and computational overhead for stronger security. 
We begin by detailing the extended adversary model and security goals (Section~\ref{sec:extension-security-considerations}), then provide a high-level overview of the protocol's design (Section~\ref{sec:extension-overview}), and finally describe the full protocol in detail (Section~\ref{sec:extension-protocol}).

\subsection{Security considerations}\label{sec:extension-security-considerations}
\point{Adversary model}
In this protocol extension, we consider a hardware adversary \hwadversary, who can bypass the protection of the TEE.
\hwadversary can access and modify keys, the protocol code and trusted node list inside the TEE. 

\point{Security goals}
Since \hwadversary has access to all keys and protocol code, it is not possible to prevent this adversary from forging an attestation report for the compromised node.
Therefore, we aim to prevent the adversary from forging attestation proofs for uncompromised nodes.
Specifically, consider a node $n$ that is not compromised by \hwadversary and for which the adversary has no attestation proof. 
This extension aims to prevent \hwadversary from deceiving any node into believing that $n$ is a trustworthy node. 
In other words, the aim is that the adversary can only convince a node that has been physically compromised that an untrusted node is indeed trustworthy, but cannot disseminate this misinformation to any other uncompromised nodes. 
This aim corresponds to a similar level of security that many collective attestation protocols provide against physical attackers \cite{kohnhauser2018salad, ambrosin2016sana, carpent2017lightweight, ibrahim2016darpa, kohnhauser2017scapi}.

\subsection{Extension overview}\label{sec:extension-overview}
To mitigate the risks posed by a \hwadversary, the protocol extension introduces cryptographic proofs that bind each attestation to its origin. 
Specifically, every node generates a verifiable proof of authenticity alongside its attestation data. 
This proof is shared with its verifier during direct attestation and subsequently propagated to other peers through trust sharing.

The goal is to ensure that when a node receives transitive trust information, it can independently verify that the underlying attestation originated from a genuine TEE and was not fabricated by a compromised or malicious party. 
By enforcing proof verification as a prerequisite for establishing transitive trust, the protocol significantly limits the ability of physically compromised nodes to inject false trust into the network.

\point{Key distribution problem}
Introducing authentication to our protocol presents a key distribution problem, which is particularly challenging to address in our peer-to-peer decentralised network setup.

Message authentication codes (MACs) are not suitable for authentication in our setting.
Since every node in our protocol can act as a verifier, it would be impractical for each node to maintain a separate secret key with every other node.
SALAD \cite{kohnhauser2018salad} uses MACs to protect against hardware attacks, but it relies on a central verifier that possesses all the necessary keys for MAC verification.
One of the core goals of our protocol is to maintain a peer-to-peer system and therefore avoid a central verifier.

Traditional digital signatures are also not feasible for authenticating attestations in our case, because they require each node to know the public key of every other node. 
Collective attestation protocols such as SANA \cite{ambrosin2016sana} and PASTA \cite{kohnhauser2019pasta} use digital signatures without addressing the key distribution problem, presumably because these protocols assume a single network owner that can distribute the public keys.
We consider a dynamic network setup, where the set of nodes joining the network is not predetermined and connections are unreliable.
Therefore, using a public key infrastructure (PKI) to distribute public keys is not feasible either, as nodes may be unable to reach the key servers for extended periods, causing delays in verifying the authenticity of attestations and even allowing a malicious attacker to launch denial-of-service attacks.
Sharing public keys between nodes is similarly problematic both due to high communication costs, the goal to be resilient to unstable connectivity, and the high memory cost of storing public keys for all trusted nodes for periodical attestation renewals (see \S\ref{sec:additional-features}).

\point{Identity-Based Signatures}
To overcome the key distribution challenge in Careful Whisper's decentralised peer-to-peer setup, we use identity-based signatures \cite{shamir1985identity} for authenticating attestations. 
In an identity-based signature scheme, a public key generator (PKG) publishes a master public key.
Participants obtain their individual private keys associated their authorised identities by contacting the PKG.
Identity-based signatures resolve the key distribution problem, since anyone who knows the master public key and the identity of any signer can verify signatures of that signer \cite{shamir1985identity}.

However, using identity-based signatures in our setup introduces other challenges. 
First, the need for a trusted PKG poses a challenge, as our network setup does not have a central owner or maintainer. 
Nonetheless, since we propose this extension for permissioned networks, it is reasonable to assume that a consortium of manufacturers or deployers can establish a trusted PKG.

Using an identity-based signature scheme also raises issues with key revocation, since a node's key cannot be revoked without changing its identity, and revoking the master key will revoke the private keys for all nodes \cite{green2015forward}.
This is an issue, because if a hardware-compromised node is detected, we would like to be able to permanently disallow that node from participating in the protocol.
To address this, we do not use node IDs as the identities for the signature scheme, but a combination of the node IDs and an epoch timestamp, which is a technique used in puncturable encryption \cite{green2015forward}. 
The PKG keeps a denylist for nodes to be removed from the network, and to remove a node, the PKG does not issue a private key for that node in the coming epochs.

The tradeoff, however, is that this approach requires nodes to periodically reconnect to the PKG to obtain updated keys.
This may be impractical in networks with unstable or intermittent connectivity.
One potential mitigation is to adjust the epoch length based on the expected threat model and desired availability.
Another option is to allow nodes to prefetch keys for upcoming epochs within a configurable time window, reducing the likelihood of disruption.

\subsection{Extension protocol}\label{sec:extension-protocol}
The first change to the core protocol described in \S\ref{sec:protocol-description} is that we define a new subprotocol \getkeyblank.
This subprotocol should be executed by each node before joining the network and also periodically before the identity-based signature private keys for each epoch expires.

\nopoint{\getkey} 
This subprotocol allows a node to request its epoch-specific private key from the PKG in order to participate in the extended protocol.
\begin{enumerate}
    \item The node sends a request to the PKG containing its node ID ($ID$), the target epoch timestamp ($\textit{epoch}$), and its certificate.
    \item The PKG verifies the node using the node's certificate; the same way that the nodes verify each other in the permissioned network.
    \item The PKG also checks if $ID$ is included in the denylist.
    \item If the node's certificate is verified and $ID$ is not in the denylist, PKG sends the node its private key for the requested $\textit{epoch}$.
\end{enumerate}

In addition to this new subprotocol, we also make two minor modifications to the \attestblank and \syncblank subprotocols as follows.

\point{Changes to \attest}
\begin{enumerate}
    \item At the end of a successful attestation in Step~\ref{item:attest-succeeds}, the verifier shares the policy it has created with the prover ($n_2$).
    \item Upon receiving the policy, $n_2$ signs it using its identity-based signature private key and sends the signature to $n_1$.
    \item $n_1$ derives a public key from the PKG's master public key using a combination of $n_2$'s node ID and the current epoch.
    \item $n_1$ verifies the validity of $n_2$'s signature using this derived public key, and stores the signature in its trusted nodes list along with $n_2$'s ID and the corresponding policy.
\end{enumerate}

\point{Changes to \sync}
\begin{enumerate}
    \item In Step~\ref{item:sync-send-list}, in addition to sharing a list of node IDs and policies with the verifier ($n_1$), the prover ($n_2$) also includes the policy signatures generated during the modified \attestblank subprotocol.
    \item In Step~\ref{item:sync-update-list}, $n_1$ verifies each received signature before adding the corresponding entries to its trusted nodes list.
    It does so by deriving a public key from the master public key using each node ID and corresponding epoch, then checking that the signature is valid for the given policy using that derived public key.
\end{enumerate}
\section{Evaluation}\label{sec:evaluation}

We evaluate Careful Whisper's performance through extensive simulation, focusing on three research questions: 
\begin{questions}
    \item How does the protocol scale with increasing network sizes and different network topologies? (\S\ref{sec:careful-resource-efficiency})
    \item How quickly and effectively does trust propagate across the network under different protocol strategies? (\S\ref{sec:careful-trust-prop})
    \item How resilient is Careful Whisper to attestation failures, and how do failure rates affect trust convergence? (\S\ref{sec:careful-failure})
\end{questions}

We address these questions through a series of discrete-event simulations (\S\ref{sec:evaluation-implementation}) under varying network topologies (\S\ref{sec:evaluation-topologies}).
We compare Careful Whisper's performance against alternative approaches (\S\ref{sec:evaluated-approaches}).

\subsection{Evaluated approaches}\label{sec:evaluated-approaches}
To understand the design tradeoffs of Careful Whisper, we compare three approaches. 
The first one is the original protocol described in \S\ref{sec:protocol}.

We also evaluate a variation of the protocol, where nodes share entire lists of trusted nodes without performing any filtering via Bloom filters. 
We include this variant to establish an upper bound on propagation speed in the absence of any memory-saving mechanism.

Finally, we include the naïve approach we used as a baseline in designing our protocol, where nodes only establish trust through direct peer-to-peer attestation and do not gossip trust information beyond what they attest themselves. 

\begin{figure*}[htbp]
  \centering
  \begin{subfigure}[t]{0.45\textwidth}
    \centering
    \includegraphics[width=\textwidth]{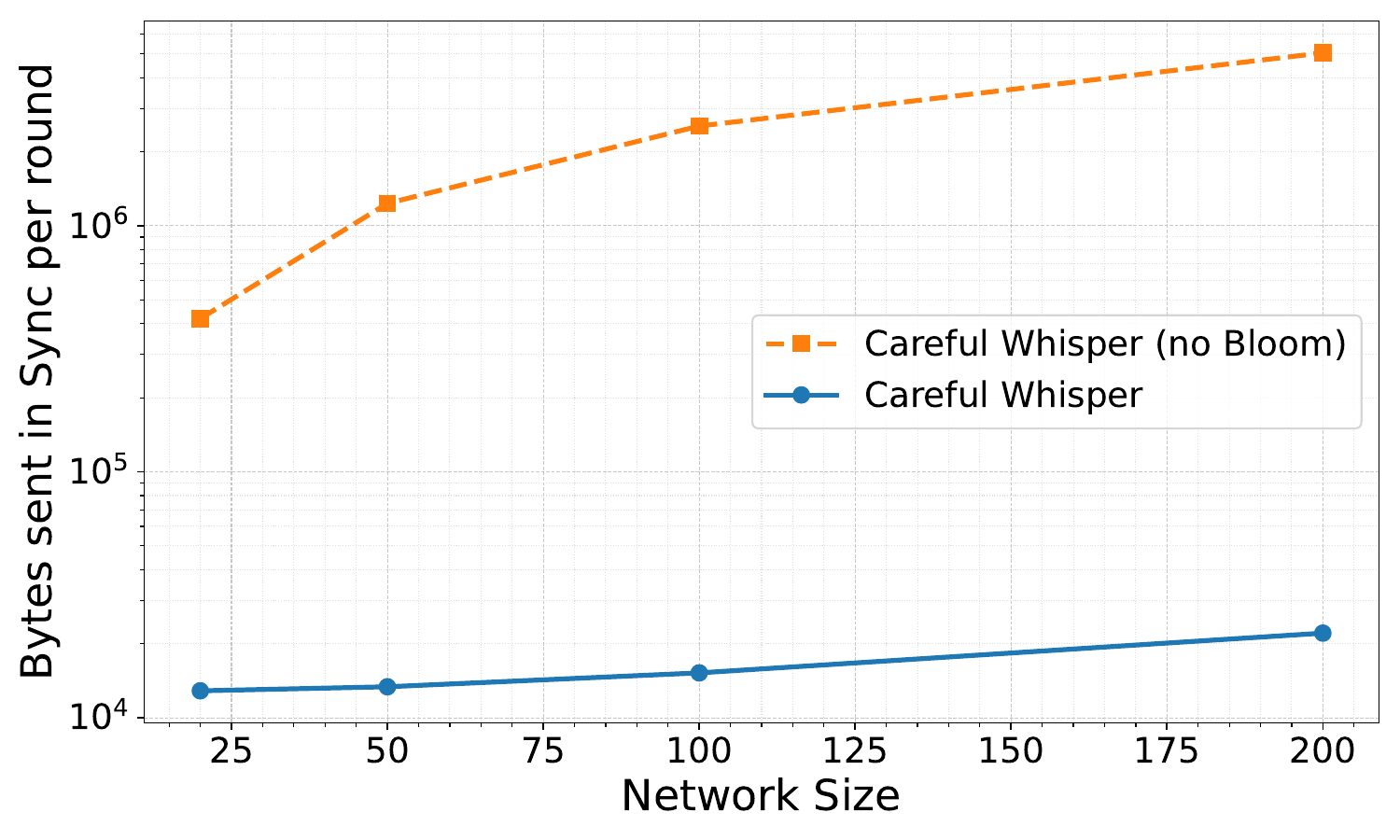}
    \caption{Average peer-to-peer message size per round.}
    \label{fig:sync_size}
  \end{subfigure}%
  \hfill
  \begin{subfigure}[t]{0.45\textwidth}
    \centering
    \includegraphics[width=\textwidth]{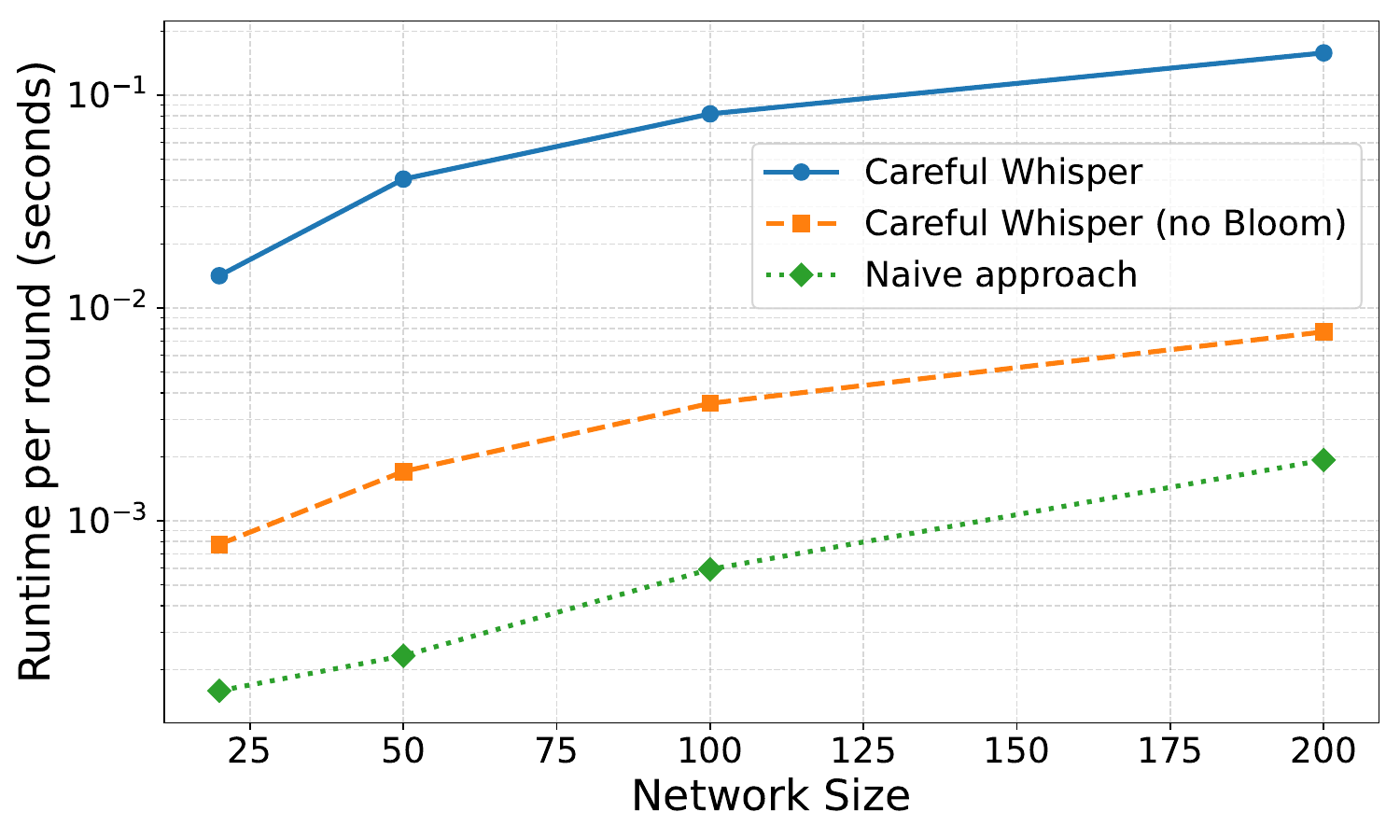}
    \caption{Average runtime per round.}
    \label{fig:runtime}
  \end{subfigure}
  \caption{Resource usage of Careful Whisper and other approaches as a function of network size.}
  \label{fig:resource-usage}
\end{figure*}

\subsection{Network topologies}\label{sec:evaluation-topologies}
To assess the effectiveness of Careful Whisper under different network conditions, we evaluate the protocol across four well-established network topologies:

\nopoint{An \textbf{Erdős--Rényi graph}~\cite{erdos1960evolution}} is a random graph, where each edge between a pair of nodes is formed independently with probability $p$. 
    This topology is useful in modelling sparse, random, distributed networks, such as ad-hoc wireless networks or partially connected peer-to-peer systems.
    
\nopoint{A \textbf{Watts--Strogatz graph}~\cite{watts1998collective}} is used for modelling small-world networks, where each node is initially connected to its \( k\) nearest neighbours, and each link is randomly rewired with probability \( p\). 
    It mimics networks with local clustering and short path lengths, such as proximity-based networks.
    
\nopoint{A \textbf{Barabási--Albert graph}~\cite{barabasi1999emergence}} models a scale-free network, where most nodes have few connections, while a small number of nodes have many. 
    In this model, nodes are added one at a time and connect to \(m\) existing nodes with a probability proportional to their degree.
    Therefore, new nodes prefer to connect to already well-connected nodes.
    With a few highly connected hubs and many low-degree nodes, this topology resembles the Internet and some social networks.
    
\nopoint{A \textbf{complete graph}} represents a fully connected network where every node has a direct link to every other node. 
    While unrealistic for large-scale systems, this topology serves as an upper-bound reference for trust propagation in ideal conditions.

We selected the following parameter values for each topology to ensure moderate connectivity and meaningful trust propagation within a limited number of simulation rounds. 
For the Erdős--Rényi model, we set the edge probability to \( p = 0.05 \), which keeps the network sparse but connected. 
In the Watts--Strogatz model, we chose \( k = 4 \) for the initial lattice degree and \( p = 0.1 \) for the rewiring probability, producing small-world graphs with high clustering and some long-range shortcuts, which balances local and global trust propagation. 
For the Barabási--Albert model, we set \( m = 2 \), meaning each new node connects to two existing nodes, generating a scale-free network with sufficient sparsity to challenge trust propagation, while still forming effective hubs. 
These values were selected empirically to reflect realistic network conditions without making the network overly dense or trivially connected.

\subsection{Implementation details}\label{sec:evaluation-implementation}
We built a custom round-based simulator in Python, using \texttt{networkx} for topology generation and modelling. 
Each simulation runs for 500 synchronous rounds, during which 100 node pairs are randomly selected (with replacement) to simulate pairwise interactions. 
These interactions may include remote attestation, optional trust list sharing, and list updates depending on the protocol variant.

To enable efficient filtering of trust entries during synchronisation, we use a Bloom filter implemented with the \texttt{bitarray} library \cite{bitarray} for compact bit storage and \texttt{mmh3} (MurmurHash3) \cite{appleby2016murmurhash} for fast, non-cryptographic hashing. 
Each Bloom filter is configured to use 512 bits (64 bytes) and three independent hash functions, each derived from \texttt{mmh3} with a different seed. 

We assume each trust entry is 128 bytes, based on a conservative breakdown of components: 8 bytes for a node ID, ~40 bytes for a policy (e.g., timestamps, flags), 64 bytes for a signature (e.g., Ed25519), and 16 bytes of encoding, padding or other optional data.
These assumptions are used to estimate the communication overhead of synchronisation in each round. 

We track three primary metrics: average trust (the average number of peers each node trusts per round),
network overhead (bytes exchanged during synchronisation), and
runtime (simulation time per round). 
These measurements allow us to compare propagation efficiency and cost across variants and conditions.

\subsection{Resource efficiency}\label{sec:careful-resource-efficiency}

\begin{figure*}[ht!]
  \centering

  \hspace{0.06\textwidth} 
  \makebox[0.3\textwidth][c]{\centering \shortstack{\scriptsize \textbf{Careful Whisper} \\ \scriptsize \textbf{Original}}}
  \makebox[0.3\textwidth][c]{\centering \shortstack{\scriptsize \textbf{Careful Whisper} \\ \scriptsize \textbf{w/o Bloom filter}}}
  \makebox[0.3\textwidth][c]{\centering \shortstack{\scriptsize \textbf{Naïve} \\ \scriptsize \textbf{Approach}}}\\[0.5em]

  \rotatebox{90}{\parbox[c]{5em}{\scriptsize \centering \textbf{Erdős} \\ \scriptsize \textbf{--Rényi}}}%
  \begin{subfigure}[b]{0.31\textwidth}
    \includegraphics[width=\textwidth]{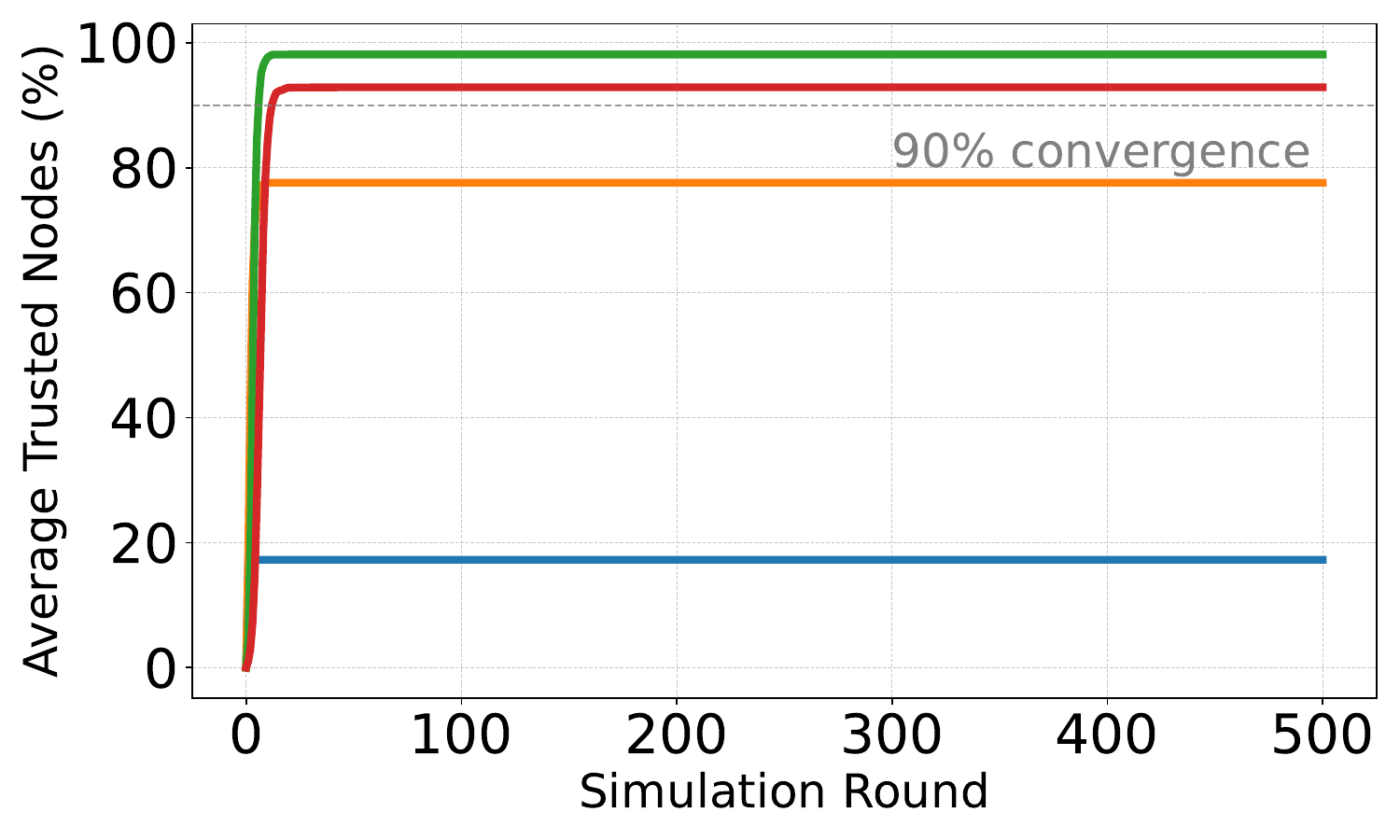}
  \end{subfigure}
  \begin{subfigure}[b]{0.31\textwidth}
    \includegraphics[width=\textwidth]{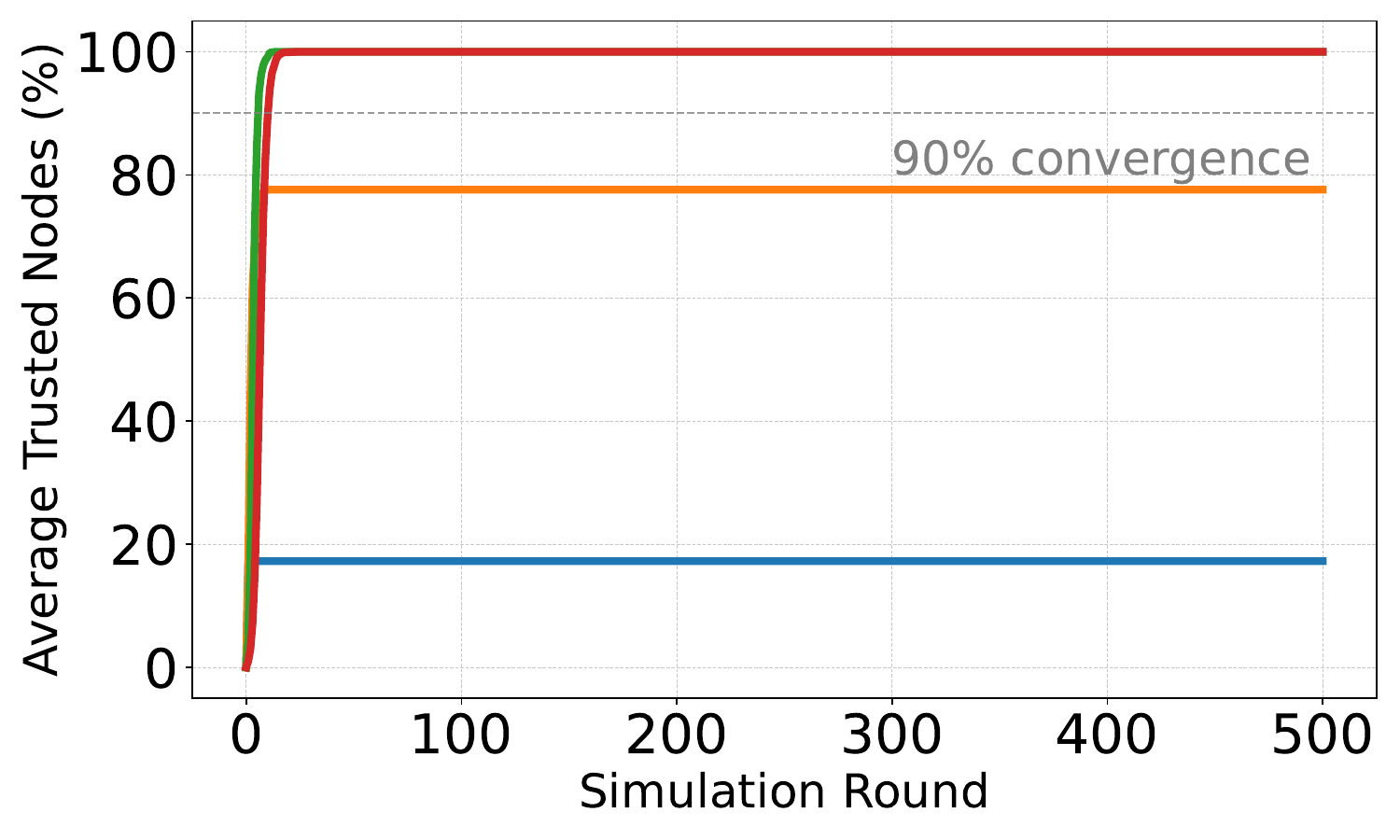}
  \end{subfigure}
  \begin{subfigure}[b]{0.31\textwidth}
    \includegraphics[width=\textwidth]{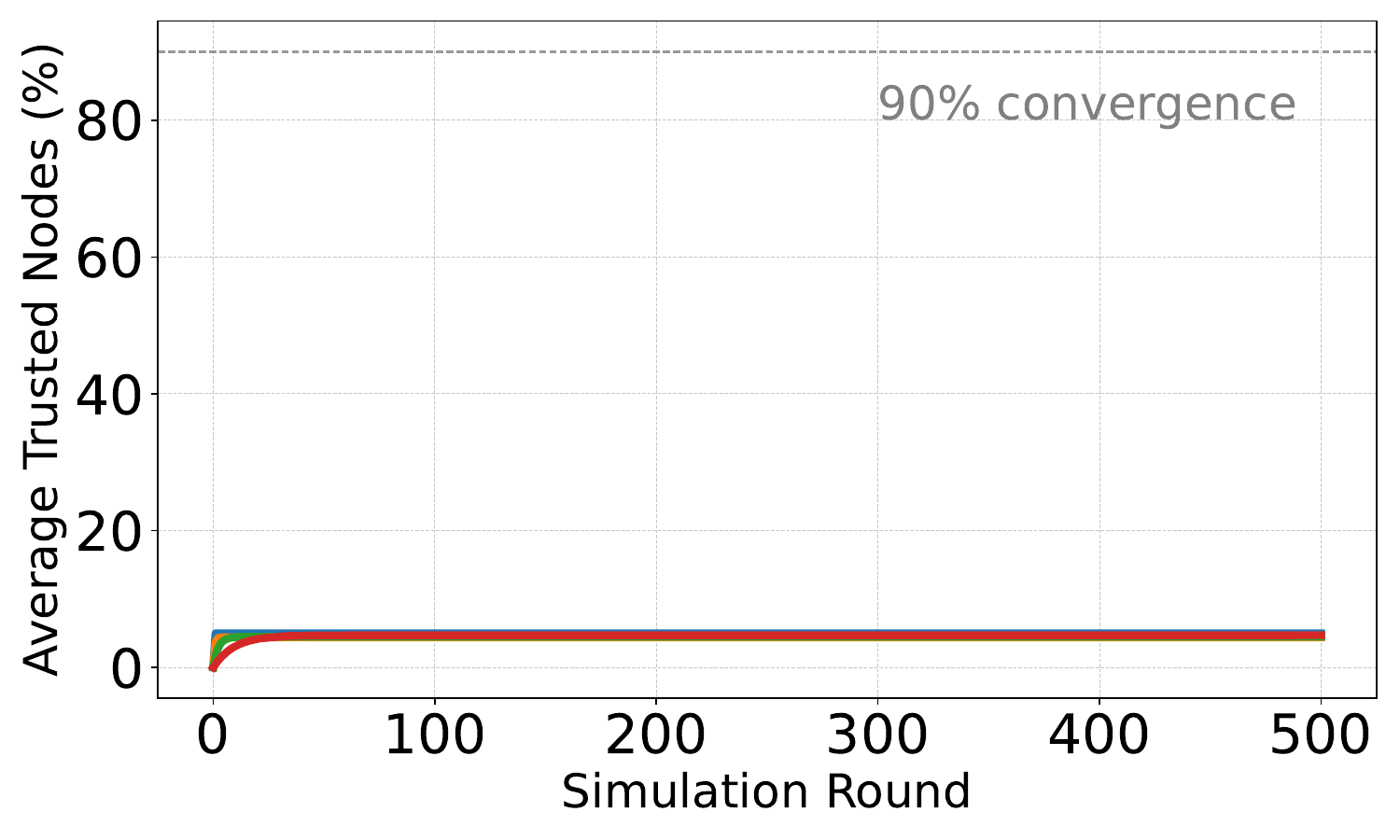}
  \end{subfigure} \\[0.75em]

  \rotatebox{90}{\parbox[c]{5em}{\scriptsize \centering \textbf{Watts--Strogatz}}}%
  \begin{subfigure}[b]{0.31\textwidth}
    \includegraphics[width=\textwidth]{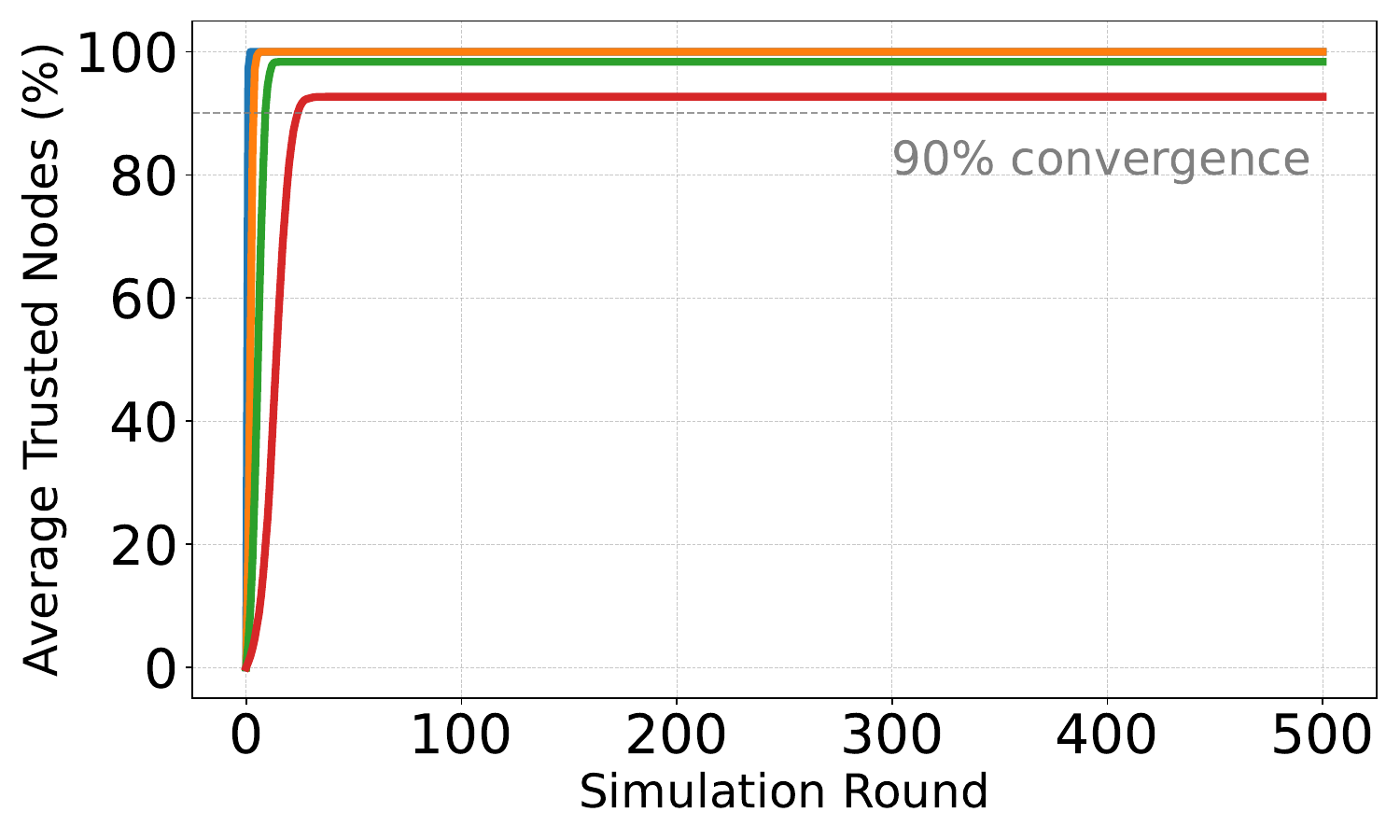}
  \end{subfigure}
  \begin{subfigure}[b]{0.31\textwidth}
    \includegraphics[width=\textwidth]{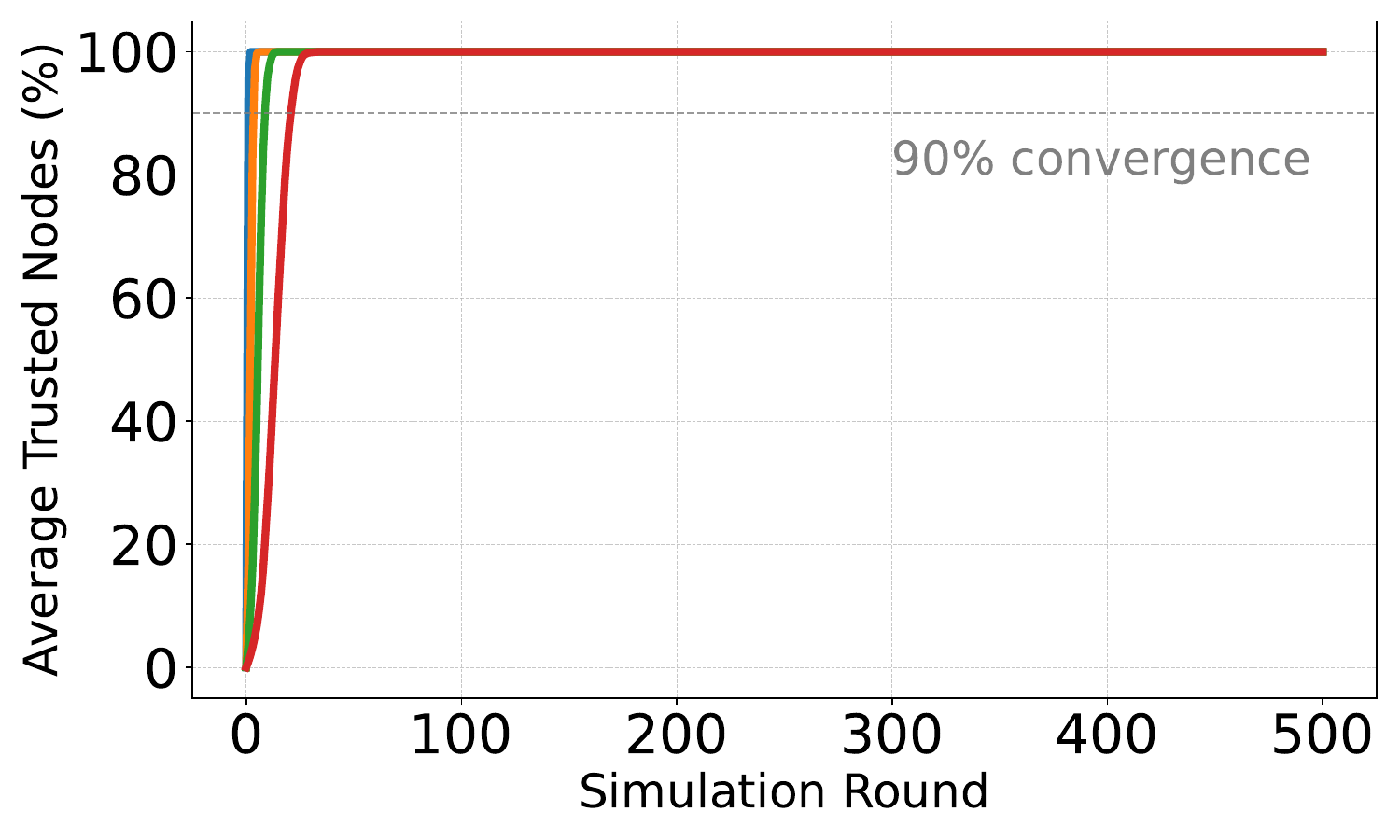}
  \end{subfigure}
  \begin{subfigure}[b]{0.31\textwidth}
    \includegraphics[width=\textwidth]{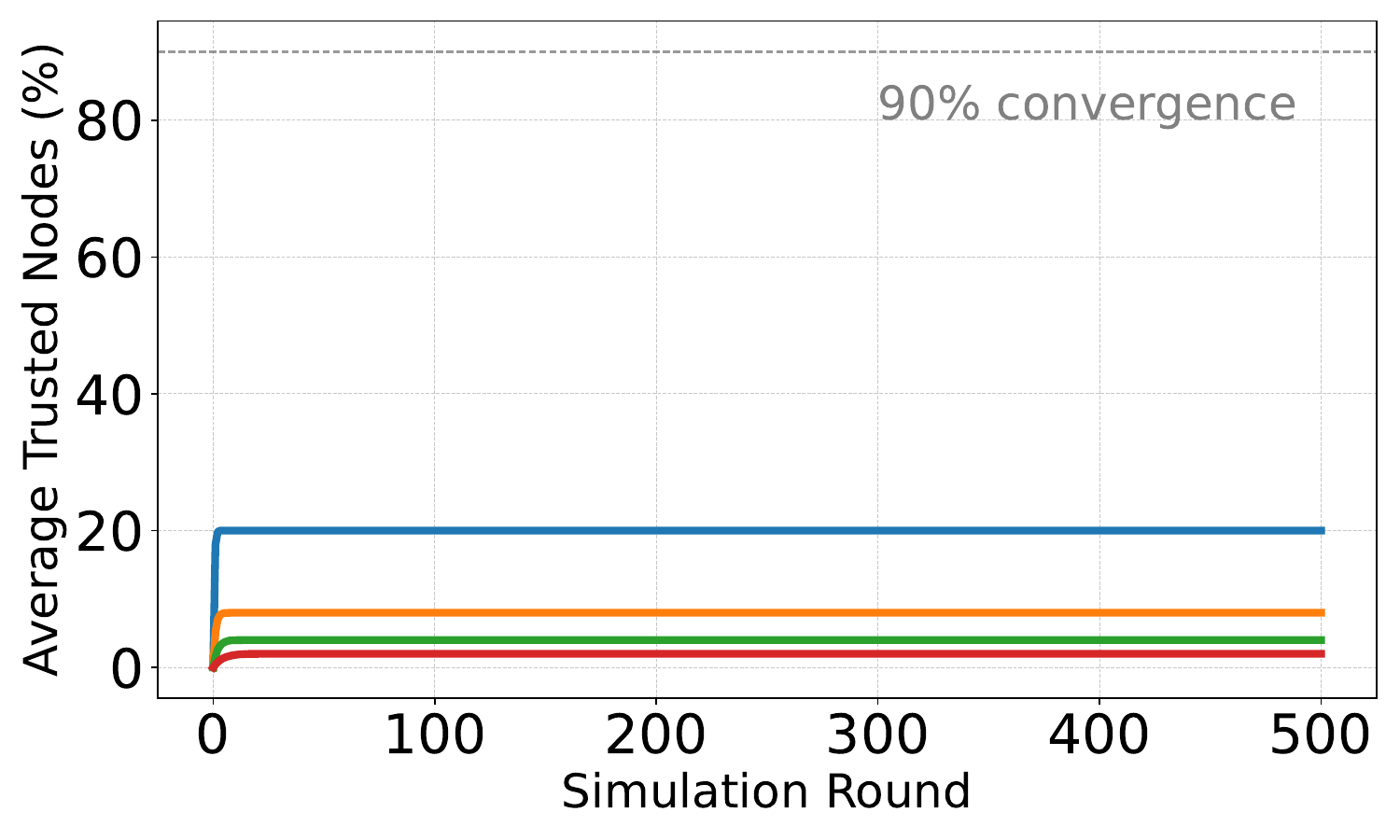}
  \end{subfigure}

  \rotatebox{90}{\parbox[c]{5em}{\scriptsize \centering \textbf{Barabási--Albert}}}%
  \begin{subfigure}[b]{0.31\textwidth}
    \includegraphics[width=\textwidth]{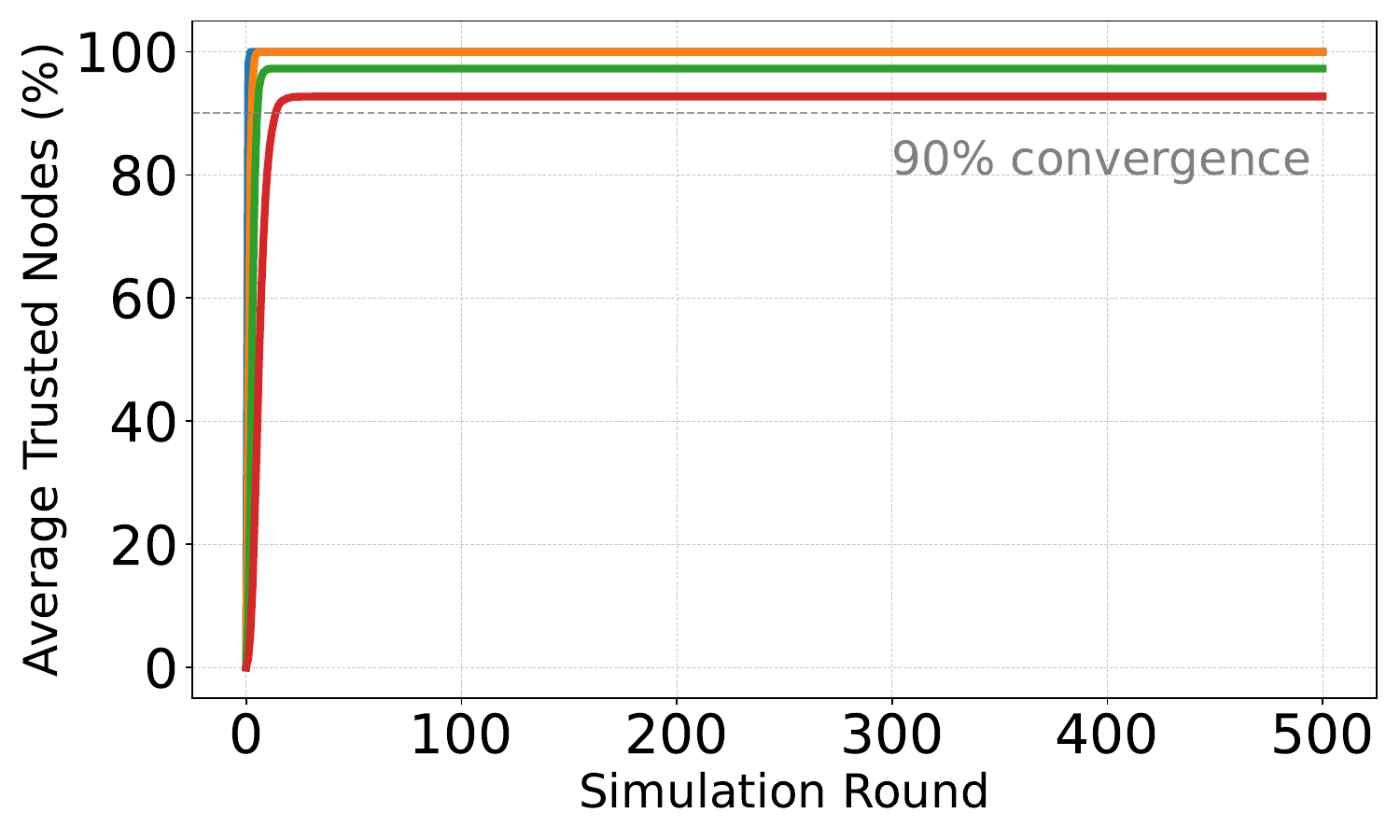}
  \end{subfigure}
  \begin{subfigure}[b]{0.31\textwidth}
    \includegraphics[width=\textwidth]{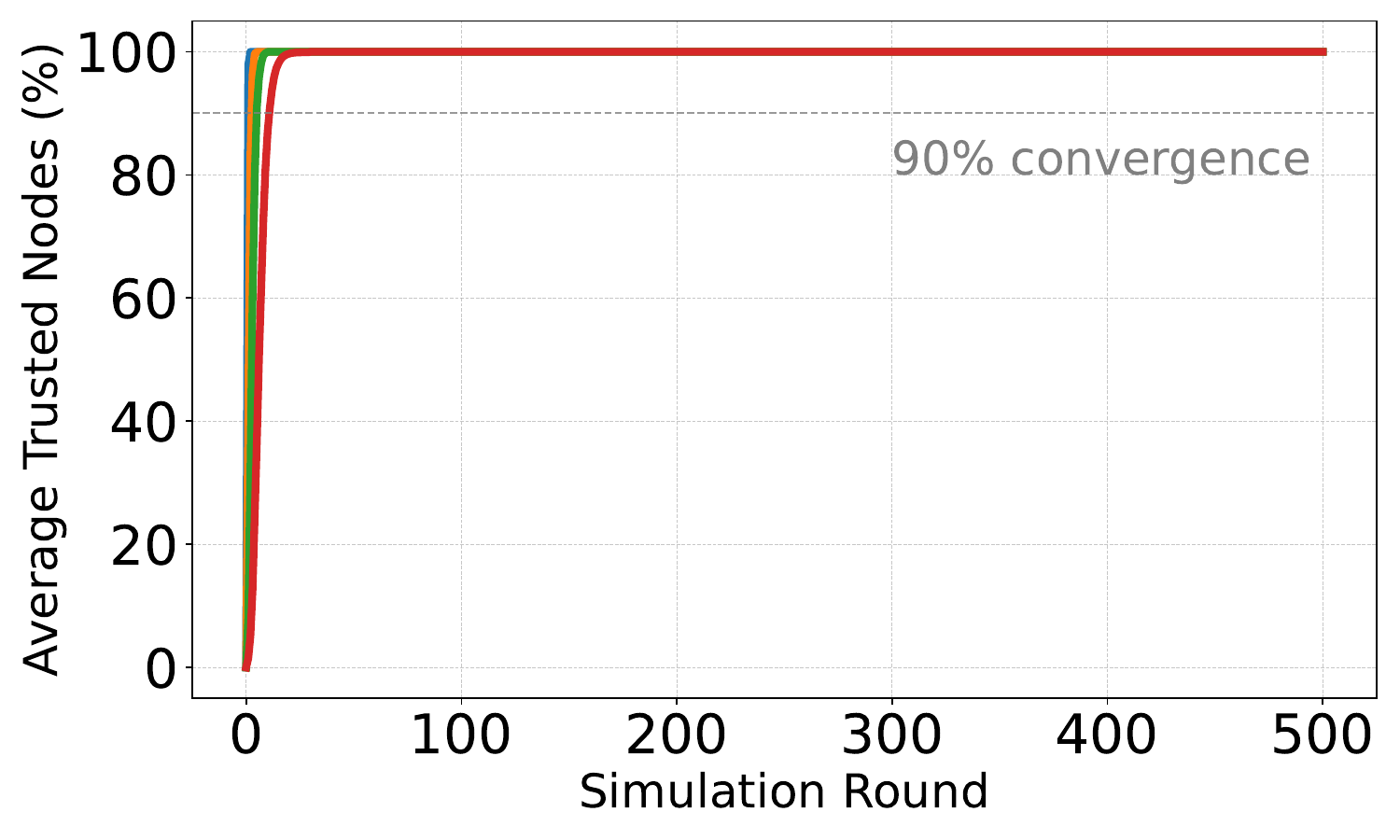}
  \end{subfigure}
  \begin{subfigure}[b]{0.31\textwidth}
    \includegraphics[width=\textwidth]{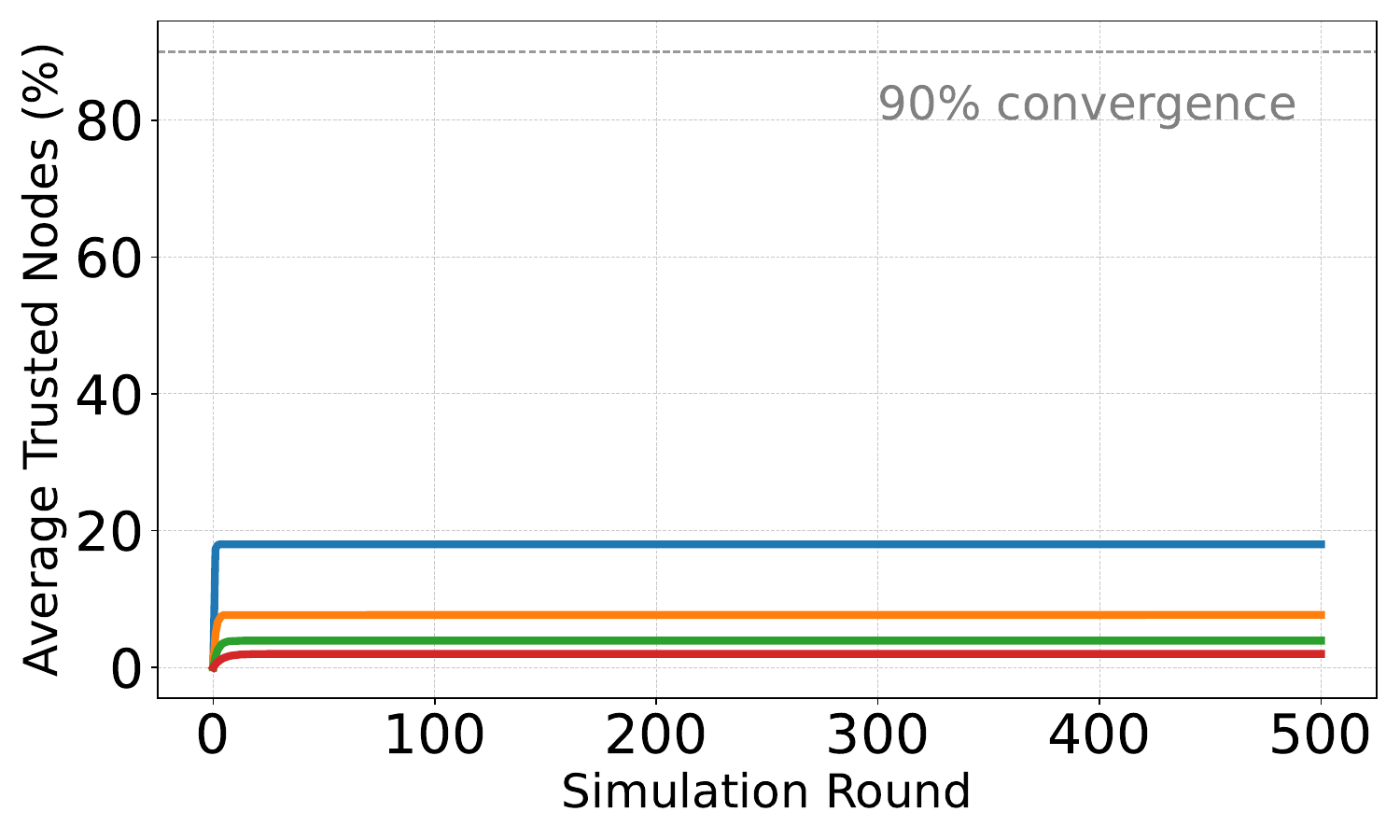}
  \end{subfigure} \\[0.75em]

  \rotatebox{90}{\parbox[c]{5em}{\scriptsize \centering \textbf{Complete graph}}}%
  \begin{subfigure}[b]{0.31\textwidth}
    \includegraphics[width=\textwidth]{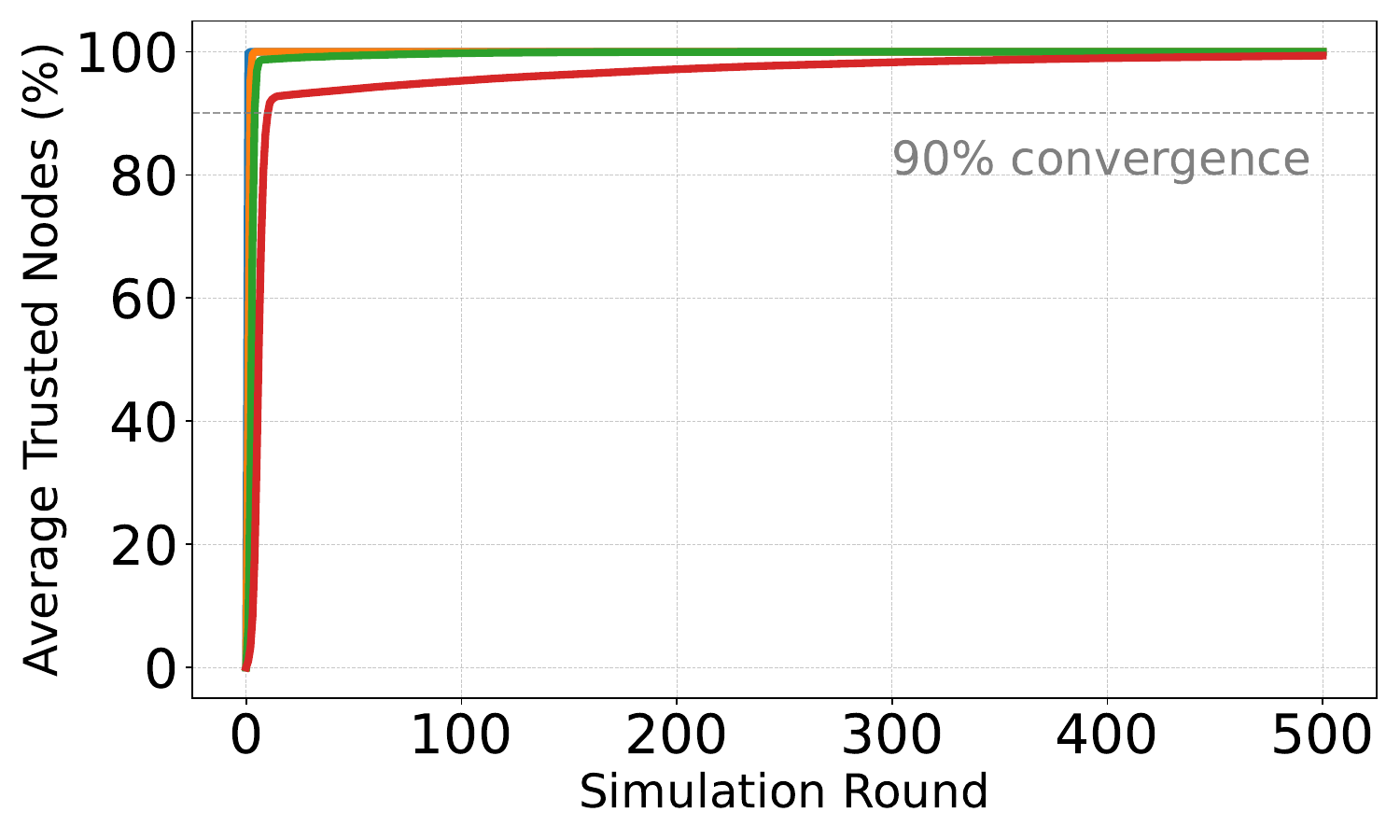}
  \end{subfigure}
  \begin{subfigure}[b]{0.31\textwidth}
    \includegraphics[width=\textwidth]{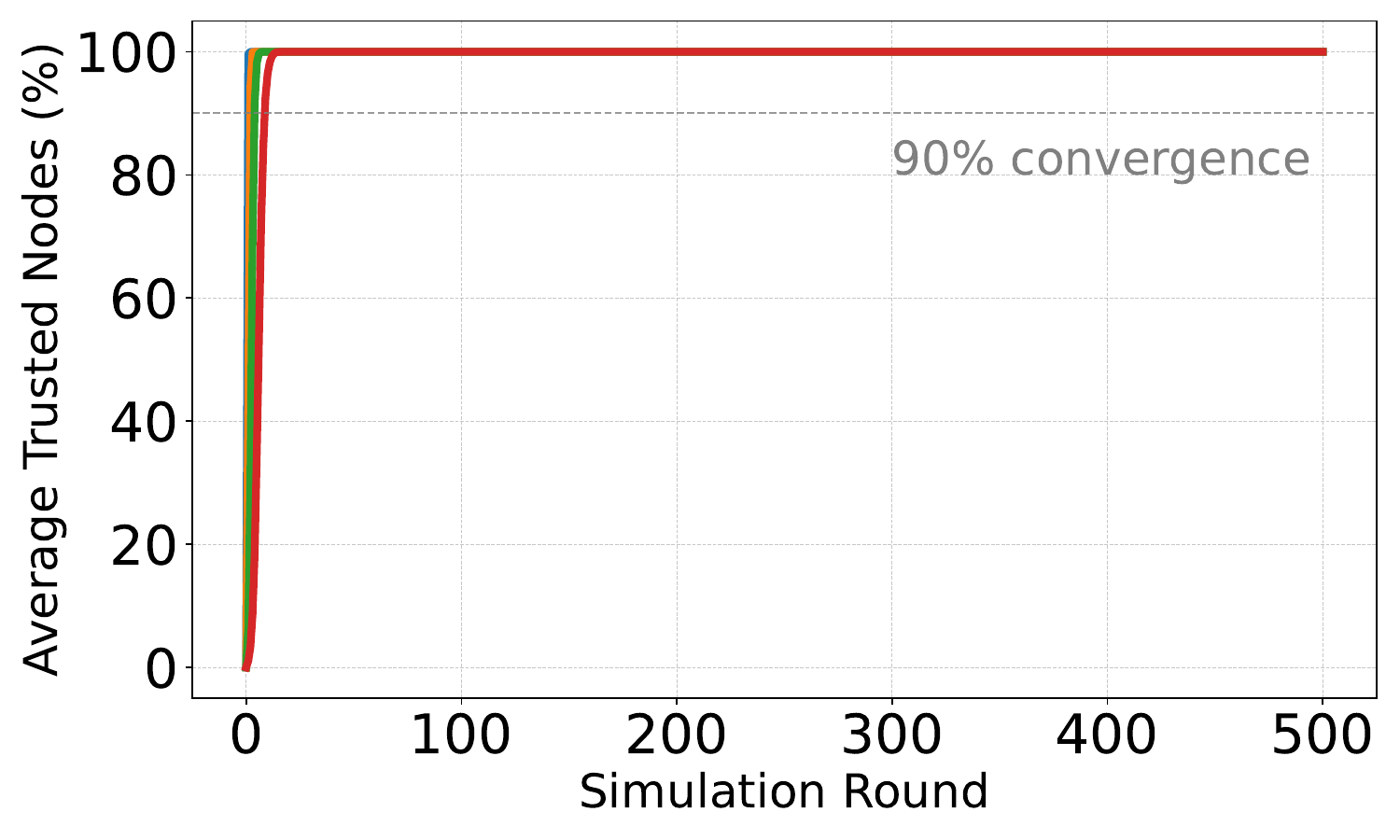}
  \end{subfigure}
  \begin{subfigure}[b]{0.31\textwidth}
    \includegraphics[width=\textwidth]{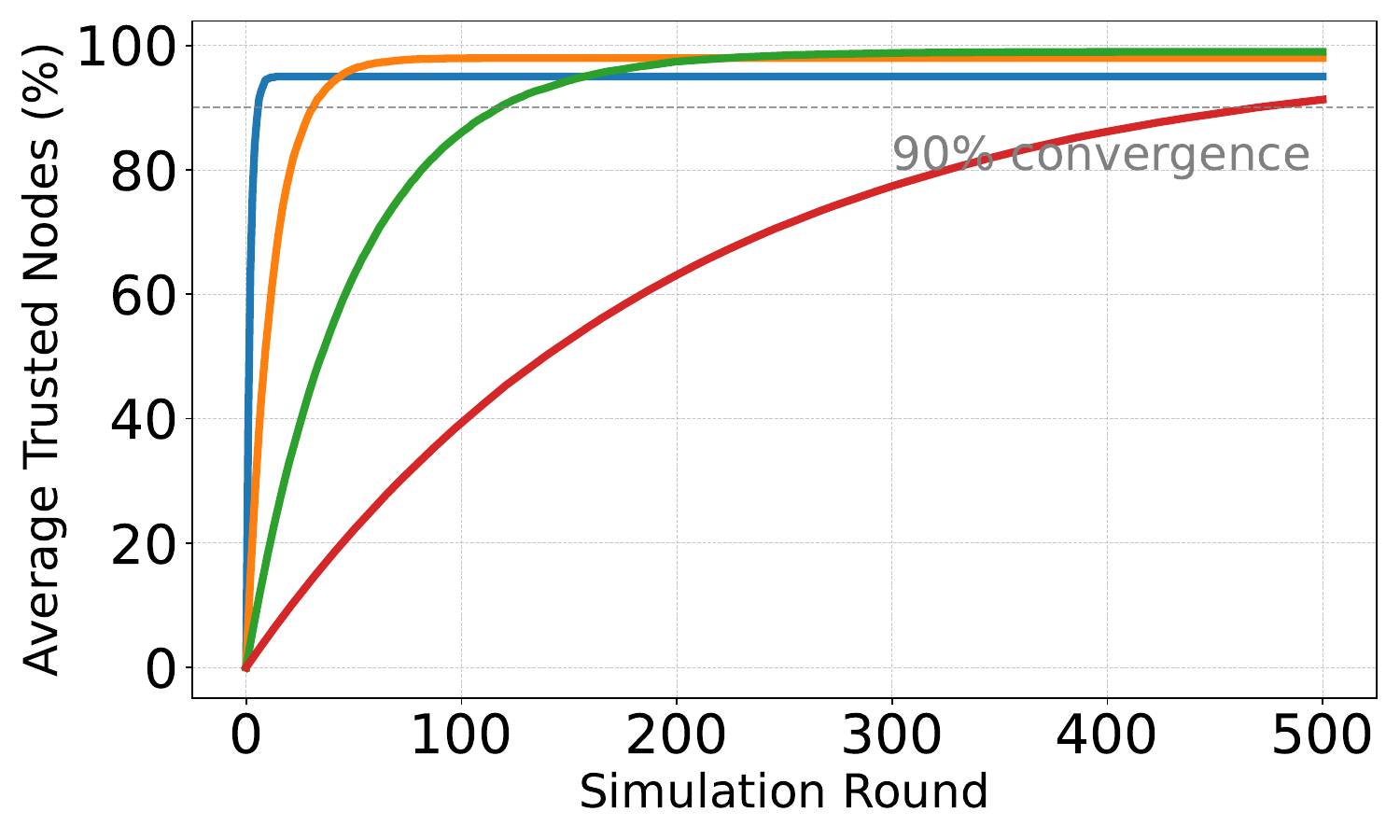}
  \end{subfigure} \\[0.75em]

{
\centering
  \begin{subfigure}[b]{0.6\textwidth}
    \includegraphics[width=\textwidth]{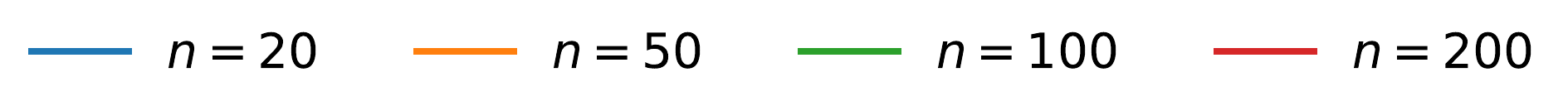}
  \end{subfigure}
}

  \caption{Trust propagation across different topologies (rows) and protocol strategies (columns).}
  \label{fig:trust_grid_rotated}
\end{figure*}

To evaluate the protocol’s efficiency, we recorded the runtime and the total number of bytes exchanged per round. 
These metrics were collected across all protocol variants and topologies, each tested at multiple network sizes ($n \in \{20, 50, 100, 200\}$) with five trials per configuration to ensure consistency.
Each trial consists of 500 rounds of pairwise node interactions.
In each round, 100 randomly selected node pairs (with replacement) interact according to the corresponding approaches' logic, performing attestation and optionally exchanging trusted node lists or Bloom filters. 
 
Figure~\ref{fig:sync_size} shows that the original Careful Whisper protocol maintains a nearly constant message size per round across different network sizes. 
In contrast, the no-Bloom-filter variant shows a linear increase in communication overhead, as nodes share their full trust lists.
This confirms the effectiveness of Bloom filters in limiting communication to only likely-missing entries.

Figure~\ref{fig:runtime} shows that while all variants experience longer per-round runtime with larger networks, the original Careful Whisper protocol consistently incurs more computational cost than the no-Bloom-filter version. 
This is because to scan a verifier's Bloom filter to identify the nodes it potentially misses in its trusted list, the verifier needs to compute multiple hashes (in our case 3) for each of its trusted nodes.
As expected, the naïve protocol has the lowest runtime, since it does not involve any steps other than attestation, but at the cost of much slower trust propagation (see Figure~\ref{fig:trust_grid_rotated}) and higher network overhead (see Figure \ref{fig:sync_size}).

\subsection{Trust propagation}\label{sec:careful-trust-prop}

To evaluate how quickly trust propagates, we simulate 500 rounds of pairwise interactions across four network sizes ($n \in \{20, 50, 100, 200\}$), with five trials per configuration. 
In each round, 100 random edges from the network are sampled (with replacement) to interact. 
After each round, we compute the average number of nodes that each node trusts.
This metric, which we refer to as \textit{average trust}, serves as a proxy for how widely trust has propagated throughout the network. 

Figure~\ref{fig:trust_grid_rotated} shows trust propagation across network topologies (rows) and protocol variants (columns). 
Careful Whisper enables rapid trust propagation across all settings, consistently approaching high trust levels. 
However, in sparser topologies like Erdős--Rényi, particularly at smaller network sizes, limited connectivity reduces the chances of nodes encountering enough trusted peers, causing trust propagation to plateau slightly below 90\%.

The no-Bloom-filter variant of the protocol performs nearly identically in terms of propagation speed, confirming that Bloom filters provide substantial bandwidth savings (as shown in Section~\ref{sec:careful-resource-efficiency}) without compromising convergence performance.
The slight difference between the original Careful Whisper protocol and the no-Bloom-filter variant is due to the tradeoff between efficiency and completeness in trust synchronisation. 
Bloom filters introduce a controlled false positive rate, meaning that a prover may incorrectly assume the verifier already knows certain trust entries and omit them during synchronisation. 
This can slightly delay the spread of trust, particularly in the early rounds and in sparse network topologies where each piece of shared information has greater influence. 
In contrast, the no-Bloom-filter variant always shares the full trust list, guaranteeing that all missing entries are delivered.
This results in marginally faster trust propagation. 
Despite this, the original Bloom-filtered version converges nearly as quickly, and offers a balance between propagation speed and bandwidth efficiency.

The naïve approach, where nodes only establish trust through direct attestation and do not propagate trust transitively, struggles to scale. 
Trust growth quickly plateaus at low levels, especially in sparser graphs, as nodes can only trust those they directly meet. 
This limitation becomes especially apparent in sparser or less connected topologies, like Erdős-–Rényi, where random pairwise interactions may not cover all node pairs even after many rounds. 
Some nodes simply never encounter each other directly, and because no indirect trust propagation occurs, their trust lists remain incomplete. 
This behaviour reflects the scalability and coverage limitations of the naïve peer-to-peer attestation approach.

Network topology influences convergence speed. 
Complete graphs converge quickly due to full connectivity, while Barabási--Albert and Watts--Strogatz topologies benefit from hubs and shortcut links. 
Despite topological differences, Careful Whisper provides robust and scalable propagation across all conditions.

\begin{figure}[ht!]
  \centering

  \begin{subfigure}[b]{0.7\columnwidth}
    \centering
    \includegraphics[width=\columnwidth]{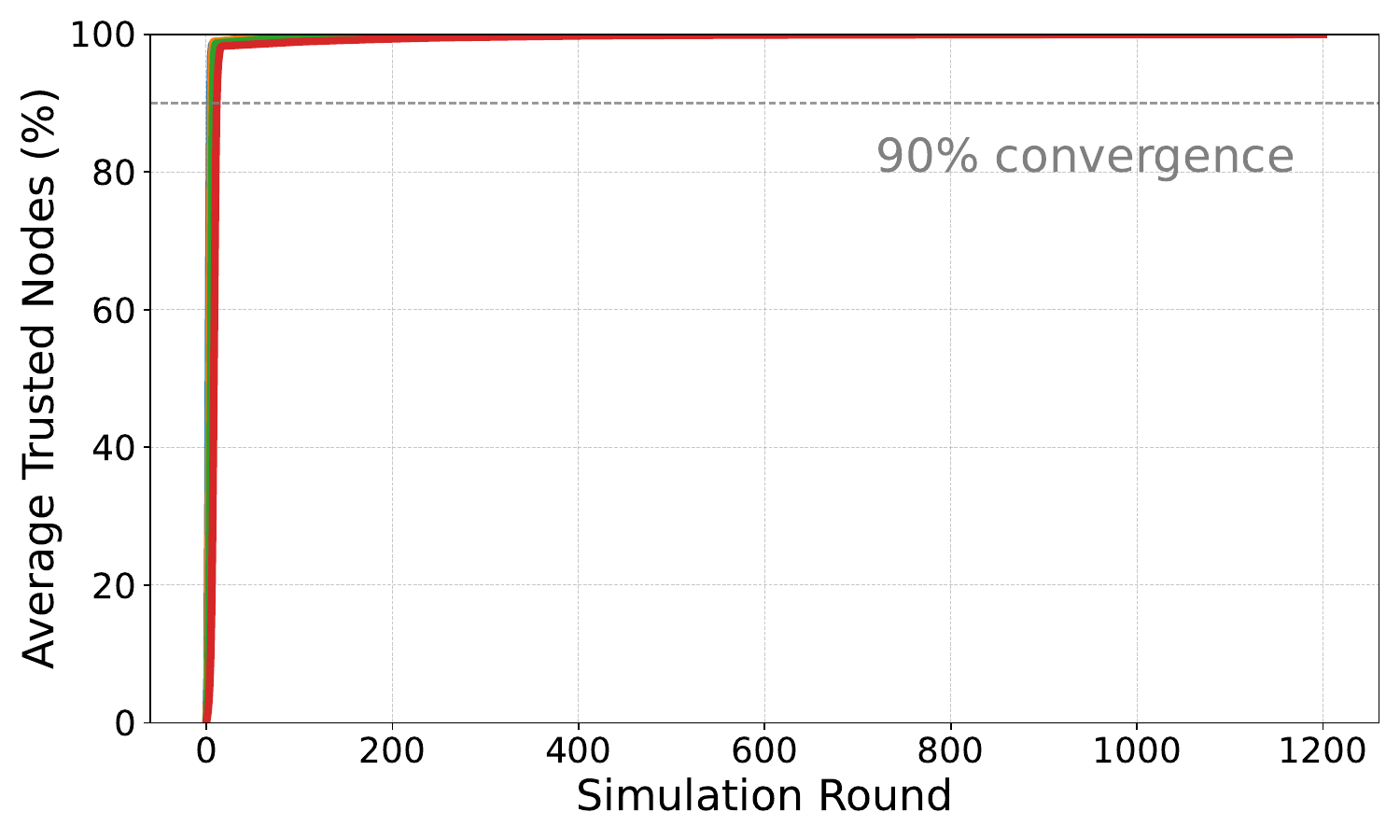}
    \caption{Careful Whisper}
  \end{subfigure}
  \vspace{1em}

  \begin{subfigure}[b]{0.7\columnwidth}
    \centering
    \includegraphics[width=\columnwidth]{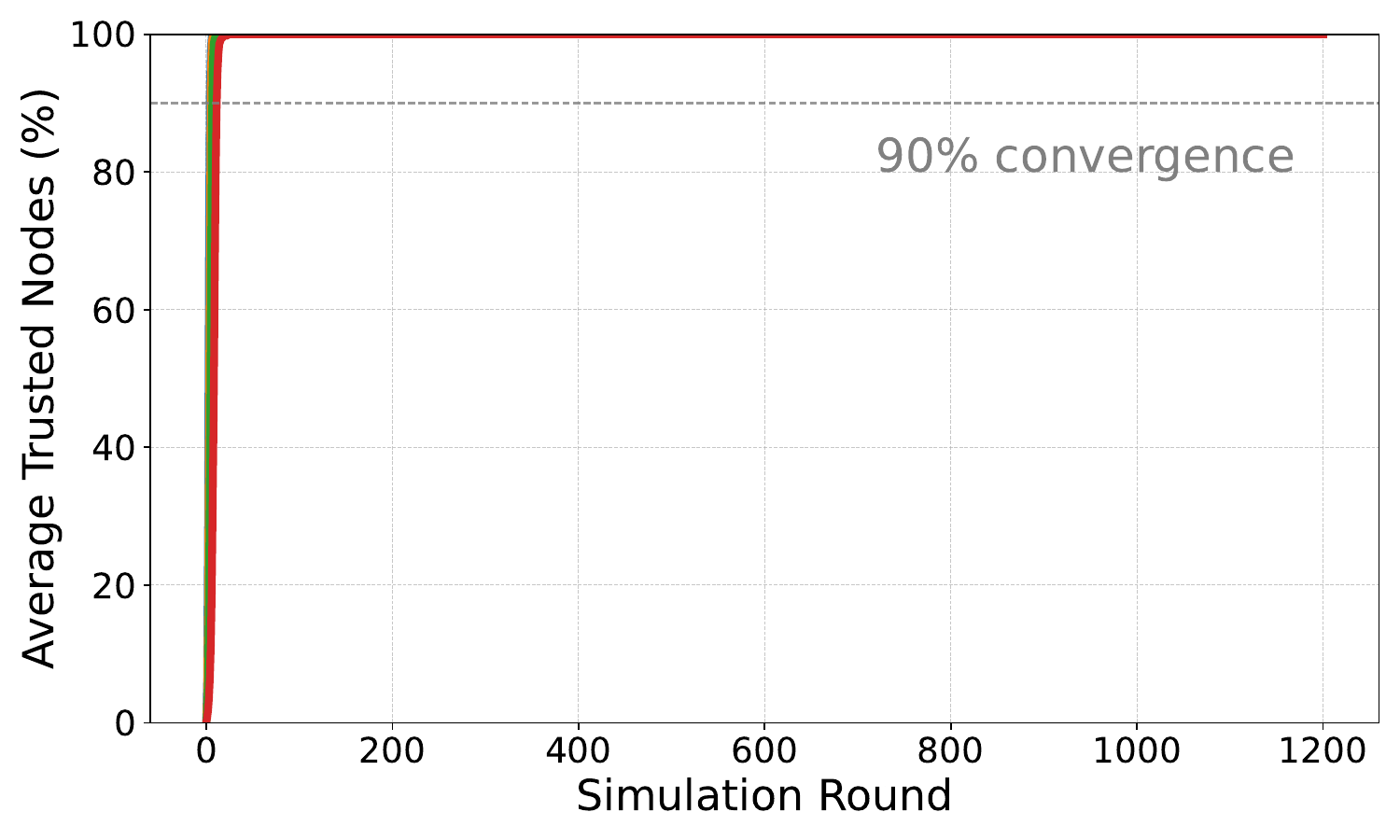}
    \caption{Careful Whisper (no Bloom filter)}
  \end{subfigure}
  \vspace{1em}

  \begin{subfigure}[b]{0.7\columnwidth}
    \centering
    \includegraphics[width=\columnwidth]{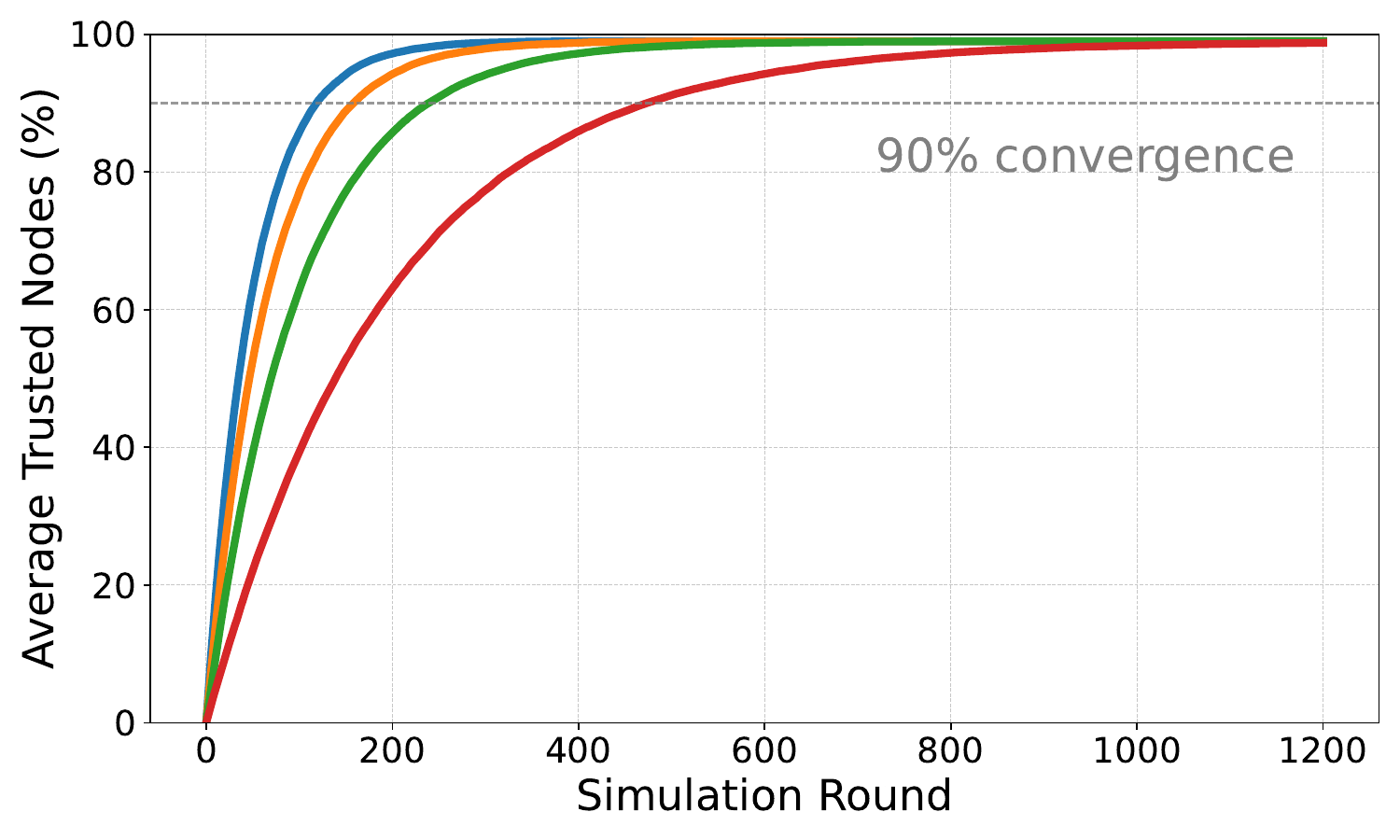}
    \caption{Naïve approach}
  \end{subfigure}

  \begin{subfigure}[b]{0.7\columnwidth}
    \centering
    \includegraphics[width=\columnwidth]{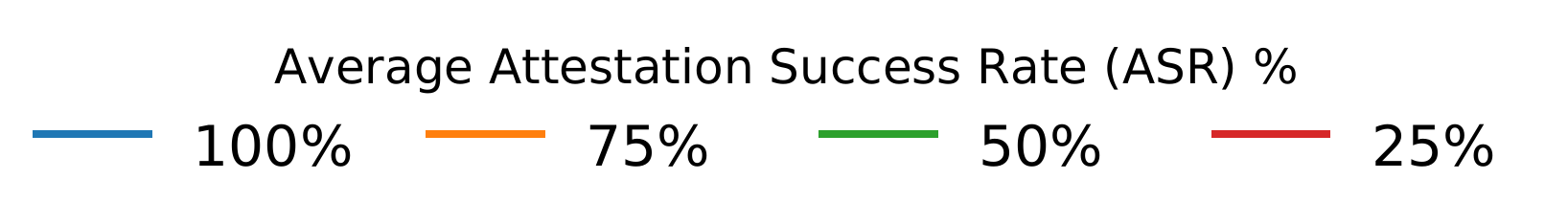}
  \end{subfigure}

  \caption{Impact of attestation failure on trust propagation in complete graph topology.}
  \label{fig:failure_resilience_complete}
\end{figure}

\subsection{Attestation failure resilience}\label{sec:careful-failure}

In real-world settings, remote attestation can intermittently fail due to a number of issues, such as network instability (e.g., dropped packets, latency), hardware issues (e.g., TEE faults or reboots), and software inconsistencies (e.g., outdated drivers, certificate mismatches). 
Service unavailability, such as downed attestation servers or reliance on third-party infrastructure, can also disrupt attestation. 
These challenges highlight the importance of designing protocols that remain robust in the face of occasional attestation failures.
We assess the protocol’s resilience by simulating a range of attestation success rates (ASR) ranging from 100\% to 25\%. 
We focus on the complete graph topology as it presents the ideal network conditions.
Figure \ref{fig:failure_resilience_complete} shows the results.

The original Careful Whisper maintains relatively high levels of trust even when attestation is unreliable. 
Trust continues to grow steadily at 75\% ASR and remains effective even at 50\% and 25\%, although more slowly. 
This shows that the protocol successfully compensates for missed attestations through gossiping with peers.
The no-Bloom-filter variant of Careful Whisper behaves similarly to the original protocol in terms of trust propagation.
As both versions rely on the same gossip-based mechanism to spread trust; the absence of Bloom filters mainly affects bandwidth efficiency rather than trust propagation under attestation failure.
On the other hand, the naïve approach struggles at moderate to high failure rates.

\subsection{Overview of findings}

Our evaluation demonstrates that Careful Whisper effectively balances scalability, trust propagation speed, and resilience under attestation failure. 
It achieves quick trust propagation while significantly reducing communication overhead, maintaining nearly constant sync sizes even as network size increases. 
Although the no-Bloom-filter variant of the protocol converges slightly faster due to sharing full trust lists, it incurs higher network overhead, confirming the value of Bloom filtering for bandwidth efficiency.

Across all tested topologies, Careful Whisper exhibits consistent performance, with minimal sensitivity to network structure. 
In contrast, the naïve approach leads to incomplete trust propagation, especially in sparse or randomly connected networks.

When attestation becomes unreliable, Careful Whisper continues to perform robustly, sustaining trust growth even at low attestation success rates. 
This resilience is due to its gossip-based trust dissemination, which helps compensate for missed direct peer-to-peer attestations. 
The naïve baseline, lacking this redundancy, quickly stagnates under failure.

Overall, the results shows that Careful Whisper provides scalable, resource-efficient, and resilient trust propagation, even in challenging network conditions.

\section{Related work}\label{sec:careful-related-work}
This section presents a comparison of Careful Whisper and leading collective attestation protocols.

In \emph{SEDA}~\cite{asokan2015seda}, the verifier initiates an attestation request that constructs a spanning tree of prover devices, with the verifier logically positioned as the root. 
Each prover in the tree attests its child nodes and forwards an aggregated attestation result to its parent. 
Ultimately, the verifier receives a single, cumulative report representing the integrity of the entire swarm.
This report includes only the total number of devices that successfully completed attestation; it does not specify which individual devices passed or failed.
While this design achieves scalability and efficiency, it relies on a stable network topology to preserve the spanning tree throughout the attestation process, a requirement that limits its applicability in highly dynamic or mobile networks. 
Furthermore, the verifier only learns the total number of trustworthy devices, but not their individual identities, which may be insufficient in scenarios that require fine-grained trust decisions or localised remediation.

\emph{DARPA} \cite{ibrahim2016darpa} is a supplementary protocol to existing single-verifier schemes, which aims to provide protection against physical attacks. 
It is based on the assumption that physical attacks typically require the targeted device to be temporarily disconnected or unreachable for a non-negligible amount of time. 
DARPA leverages this by implementing a heartbeat protocol: devices periodically broadcast authenticated heartbeats to prove their presence, and log the heartbeats of their peers. 
During the next attestation round, a central verifier collects these logs to detect any devices that were absent for longer than a defined threshold. 
However, DARPA assumes a centralised deployment model and still depends on a single trusted verifier for attestation coordination and final judgement. 
While it improves security in the presence of physical attacks, its reliance on centralised control and static heartbeat intervals limits its applicability in dynamic, decentralised environments.

\emph{SANA}~\cite{ambrosin2016sana} introduces a novel Optimistic Aggregate Signature (OAS) scheme that enables publicly verifiable aggregate attestation reports with constant-time verification, regardless of network size. 
To mitigate denial-of-service (DoS) attacks, SANA incorporates an authorisation token mechanism that limits which verifiers can initiate the protocol. 
SANA provides granular visibility into which devices passed or failed attestation, including their software configurations. 
Despite its advantages, SANA shares a key limitation with SEDA: it requires full device connectivity during attestation and relies on the construction of a stable aggregation tree. 
Moreover, the OAS cryptographic operations impose significant computational overhead on resource-constrained devices, which can hinder its deployment in real-world IoT environments with strict energy and processing constraints.
SANA also relies on a central verifier and assumes uniform attestation mechanisms, making it less suitable for decentralised or heterogeneous environments where nodes must establish trust autonomously.

\emph{DIAT}~\cite{abera2019diat} considers a setting where each prover is also a verifier. 
It uses a decentralised attestation mechanism, where each device is responsible for verifying the integrity of others based on their interactions. 
Although this does not require a device to directly communicate with every other node it intends to verify, the complexity of verification remains quadratic. 
This is because devices must still individually validate attestations tied to each device, and attestation results are not globally aggregated or reused.
Compared to Careful Whisper, DIAT offers strong guarantees for data integrity in collaborative settings but lacks decentralised trust propagation and flexibility across heterogeneous TEE environments.

\emph{PASTA}~\cite{kohnhauser2019pasta} uses tokens created with Schnorr multisignatures to achieve a decentralised, robust, and secure attestation scheme. 
In PASTA, provers do not exchange detailed attestation reports with one another. 
Instead, each device performs a self-attestation inside its TEE. 
If this self-attestation succeeds, each device contributes to the generation of a collective, unforgeable token.
This token includes a timestamp, a list of participating device IDs, and a joint signature over this information. 
It does not contain any explicit evidence of the device's software configuration or runtime state.
Devices can then assess the integrity of their peers by checking whether they have recently contributed to a valid token.
Nevertheless, PASTA offers limited transparency and flexibility: it does not support heterogeneous attestation methods, nor does it allow fine-grained trust relationships or evidence inspection.

\emph{SALAD}~\cite{kohnhauser2018salad} is a collective attestation protocol that uses gossiping as its core mechanism, like Careful Whisper.
However, SALAD significantly differs from our protocol in terms of its trust model, deployment assumptions, and attestation flow.
SALAD relies on a central network operator to initiate attestation rounds, define valid software states, and ultimately verify the aggregated attestation results. 
In contrast, Careful Whisper does not rely on a single verifier; it supports \textit{ad hoc} deployments where nodes can join without any central coordination, allowing for multiple or even no operators. 
SALAD's attestation proofs are based on MACs, and while the paper does not detail key distribution, it implies a pre-established trust through keys issued by the network operator during deployment. 
SALAD enforces a global definition of ``healthy'' software states determined by the operator, whereas Careful Whisper allows each device to evaluate attestation results according to its own locally defined trust policies.

In addition to its centralised nature, SALAD lacks support for heterogeneous attestation schemes and does not provide transitive trust relationships.
Moreover, SALAD focuses on maintaining trust in a network that has already been deployed, while Careful Whisper is concerned with establishing trust from an untrusted starting point. 
Careful Whisper supports both interactive and non-interactive remote attestation protocols, accommodating diverse TEEs and attestation mechanisms, whereas SALAD is effectively limited to non-interactive protocols. 

\section{Conclusions}\label{sec:careful-summary}
In this paper, we presented Careful Whisper, a protocol for scalable, decentralised trust establishment in peer-to-peer networks of trusted execution environments (TEEs). 
Careful Whisper enables nodes to gradually build indirect trust through authenticated interactions and efficient gossip-based synchronisation, without requiring central infrastructure, global views of the network, or stable connectivity.

We evaluated our protocol across a wide range of network conditions. 
Simulations showed that the protocol propagates trust rapidly, even in large and dynamic networks, while significantly reducing communication overhead by using Bloom filters. 
It remains effective under high attestation failure rates due to redundancy of gossiping. 
Compared to both a naive baseline and a variant without Bloom filters, Careful Whisper offers strong tradeoffs between propagation speed, bandwidth efficiency, and robustness, demonstrating its suitability for real-world deployment in heterogeneous, resource-constrained systems.

\bibliographystyle{ACM-Reference-Format}
\bibliography{references}

\end{document}